\documentclass[authoryear,12pt]{elsarticle}

\usepackage{amssymb}
\usepackage{eurosym}
\usepackage{amsmath}
\usepackage{amsfonts}
\usepackage{graphicx}
\usepackage{subfigure}
\usepackage[dvipsnames]{xcolor}
\journal{International Journal of Plasticity, accepted for publication.}
\begin{document}

\begin{frontmatter}

\title{Experimental and numerical analysis of cyclic deformation and fatigue behavior of a Mg-RE alloy}
\author[label1]{Meijuan Zhang}
\author[label1]{Hong Zhang}
\author[label1,cor]{Anxin Ma}
\ead{anxin.ma@imdea.org}
\author[label1,label2]{Javier Llorca}
\address[label1]{IMDEA Materials Institute \\ c/ Eric Kandel 2, 28906 Getafe, Madrid, Spain.}
\address[label2]{Department of Materials Science, Polytechnic University of Madrid \\ E. T. S. de Ingenieros de Caminos, Ciudad Universitaria, 28040 Madrid, Spain.}
\cortext[cor]{Corresponding author. tel +34 91 549 34 22}

\begin{abstract}
Strain-controlled fatigue of Mg-1Mn-0.5Nd (wt.\%) alloy were studied by experiment and simulation. The microstructure was made up of a dispersion of strongly texture grains (13\%) embedded in a matrix of grains with a random texture. The cyclic stress-strain curves showed limited tension-compression anisotropy because of the limited texture. Cyclic hardening under compression and cyclic softening under tension occurred due to the presence of twinning. Moreover, the twin volume fraction of the broken samples depended on whether the sample was broken in tension or compression, indicating that twining-detwinning occurs during the whole fatigue life. The mechanical response of the polycrystalline alloy was simulated by means of computational homogenization. The behavior of the Mg grains was modelled using a phenomenological crystal plasticity model that accounted for basal, prismatic and pyramidal slip (including isotropic and kinematic hardening) as well as twining and detwinning. The model parameters were calibrated from the cyclic stress-strain curves at different cyclic strain amplitudes. Numerical simulations were used to understand the dominant deformation mechanisms and to predict the fatigue life by means of a fatigue indicator parameter based on the accumulated plastic shear strain in each fatigue cycle.
\end{abstract}

\begin{keyword}
A.~magnesiumm \sep A.~rare earth \sep A.~fatigue \sep B.~crystal plasticity \sep B.~fatigue indicator parameter \sep slip \sep twin \sep back stress \end{keyword}

\end{frontmatter}

\section{Introduction}

Mg alloys present low density, good castability, high specific stiffness and
reasonable cost and they are being considered for different structural
applications in transportation (aerospace, automotive) 
\citep{Mordike2001, LWH20} as well as in health care (biodegradable implants) due to their
excellent biocompatibility \citep{ZGW14, Echeverry-Rendon2019}. These
driving forces have impulsed the investigation on the deformation and
fracture mechanisms of Mg alloys to improve the mechanical properties 
\citep{AN10, N12}. Mg has a HCP lattice with $c/a$ = 1.624 which leads to a
negligible critical resolved shear stress (CRSS) for $<a>$ basal slip that
cannot be easily increased through either solution \citep{WLA19} or
precipitation hardening \citep{CCP19, AL20}. Moreover, the CRSS for $<a+c>$
pyramidal slip is very high in most Mg alloys and plastic deformation along
the $c$ axis has to be accommodated by twinning, leading to a strong plastic
anisotropy that limits the ductility \citep{LP17}. So far, the Mg alloys
with best properties in terms of strength, ductility and limited plastic
anisotropy have been obtained through the addition of rare earths (Y, Gd,
Nd, Ce) which enhance pyramidal slip and reduce the strong basal texture
generated during deformation processing \citep{BNS07, SB08}.

While the relationship between microstructure and mechanical properties of
Mg and Mg alloys under monotonic deformation has been analyzed extensively
in recent years, the information about the mechanical behavior under cyclic
deformation is more limited. \cite{OS90} summarized the data
available in the literature up to 1990, which was focused in the experimental
determination of the S-N curves and fatigue crack growth rates under
different loading conditions. More recent investigations in strongly
textured Mg-Zn and Mg-Al alloy subjected to fully-reversed cyclic
deformation showed that these alloys presented
a strong tension-compression asymmetry due to
continuous activation of twinning-detwinning in grains suitably oriented 
\citep{WJB08, YZJ11, Dong2014, Xiong2014}. The strength in compression was
controlled by the stress required to activate twinning, while the strength
in tension was determined by the harder, non-basal slip mechanisms. The
tension-compression asymmetry was smaller at low cyclic strain amplitudes
because twinning was limited. Most of the twins formed during compression
were removed when the load was reversed but the residual twin volume
fraction gradually increased with the number of cycles. Moreover, {\it in situ}
cyclic deformation tests have been carried out in synchrotron beams lines to
determine accurately the critical stress for twinning an detwinning %
\citep{ZJS19, MPB19} and this information has been used within the framework
of crystal plasticity finite element simulations to predict the cyclic
stress-strain behavior \citep{ZJS19, Briffod2019}. They showed that the information provided by the simulations is
critical to assess the dominant deformation mechanisms in terms of slip and twinning which will determine the fatigue
life of Mg alloys.

Previous investigations were focused in strongly textured Mg alloys and there is very limited experiment-simulation coupled studies on low cyclic fatigue of Mg-RE alloys. The cyclic deformation behavior of a Mg-Gd-Y alloy was studied by \citep{FWang2013}, who reported the cyclic stress-strain curves and the fatigue life as a function of the cyclic strain amplitude. In a follow-up paper, \cite{FWang2014} reported from fractographic observations the mechanisms of microcrack initiation during fatigue. They found that that microcracks were nucleated at grain boundaries at high strain amplitudes and along persistent slip bands at low strain amplitudes. However, their results were purely phenomenological and did not analyze the dominant deformation mechanisms neither included any theoretical or numerical modelling. \cite{Zhu2014} studied the fatigue deformation of Mg-RE alloys experimentally and numerically, while linear fracture mechanics model was used to study fatigue life and microstructure was not considered. \cite{Chen2017} reported experimental result of fatigue of Mg-RE alloys but they did not make modelling and simulation. In this investigation, the mechanical properties
under monotonic (tension and compression) and fully-reversed cyclic
deformation were determined in an extruded Mg - 1Mn - 0.5Nd (wt. \%) alloy
along the extrusion direction. The fatigue life was determined as a function
of the cyclic strain amplitude and the deformation mechanisms were analyzed by means of computational homogenization using a phenomenological crystal-plasticity model. Finally, the fatigue life as a function of the cyclic strain amplitude was predicted from a fatigue indicator parameter based on the accumulated plastic shear strain in each slip system in each fatigue cycle

\section{Material and Experimental Techniques}

The Mg alloy containing 1 wt. \% of Mn and 0.5 wt. \% of Nd was manufactured
by gravity casting, followed by homogenization at 623 K during 15 hours and
extrusion at 573 K and 8.3 mm/s to produce round bars of 17 mm in diameter.
The extrusion ratio was 1:30 and more details can be found in \cite
{Paloma2013}. Longitudinal sections of the bars were prepared using standard
techniques and analyzed by electron backscatter diffraction (EBSD) to
determine the grain size and shape in a dual-beam field emission gun
scanning electron microscope (Helios Nanolab 600i FEI) equipped with an
Oxford-HKL electron back scattered system. In addition, the texture was
measured by means of X-ray diffraction in the central region of the bars in
sections perpendicular to the extrusion direction. The (0001), $(10\bar{1}3)$, $(10\bar{1}2)$, $(10\bar{1}1)$, $(10\bar{1}0)$ and $(11\bar{2}0)$ pole
figures were measured using Cu K$_{\alpha}$ radiation in an Empyrean
Panalytical diffractometer.

\begin{figure}[tbp]
\centering
\includegraphics[width=0.8\textwidth]{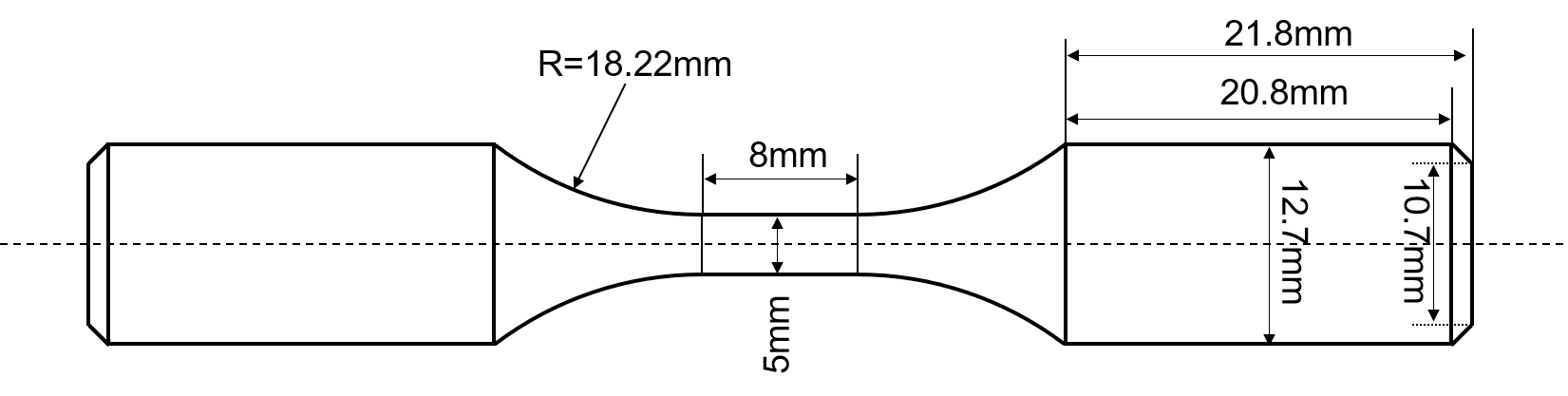}
\caption{Dimensions of cylindrical samples for the monotonic and fatigue
tests.}
\label{fig:sampleDimension}
\end{figure}

Monotonic (tension and compression) and fully-reversed, constant strain
amplitude fatigue tests were carried out in cylindrical samples machined
parallel to the extrusion direction. The dimensions of the samples are
depicted in Fig. \ref{fig:sampleDimension}. All the mechanical tests were
carried out in a servo-hydraulic mechanical testing machine. The deformation in the central section of the cylindrical specimens was
measured (and controlled in the case of the cyclic tests) with an
extensometer. Monotonic tests were carried out at an approximate strain rate
of 10$^{-3}$ s$^{-1}$ under stroke control. Fatigue tests were carried out
at three different cyclic strain semi-amplitudes of $\Delta \epsilon/2$ =
0.8\%, 2.0 \% and 4.0\% at constant strain rate of 0.001 s$^{-1}$

The specimens were initially deformed in tension in some cases and in
compression in other cases. The load and the deformation were recorded
during the tests, which finished when the sample failed. Longitudinal
sections were prepared from the fractured specimens, polished and analyzed
by EBSD at different distances from the fracture surface to determine the
fraction of twinned material.

\section{Experimental results}

\begin{figure}[tbp]
\centering
\includegraphics[width=\textwidth]{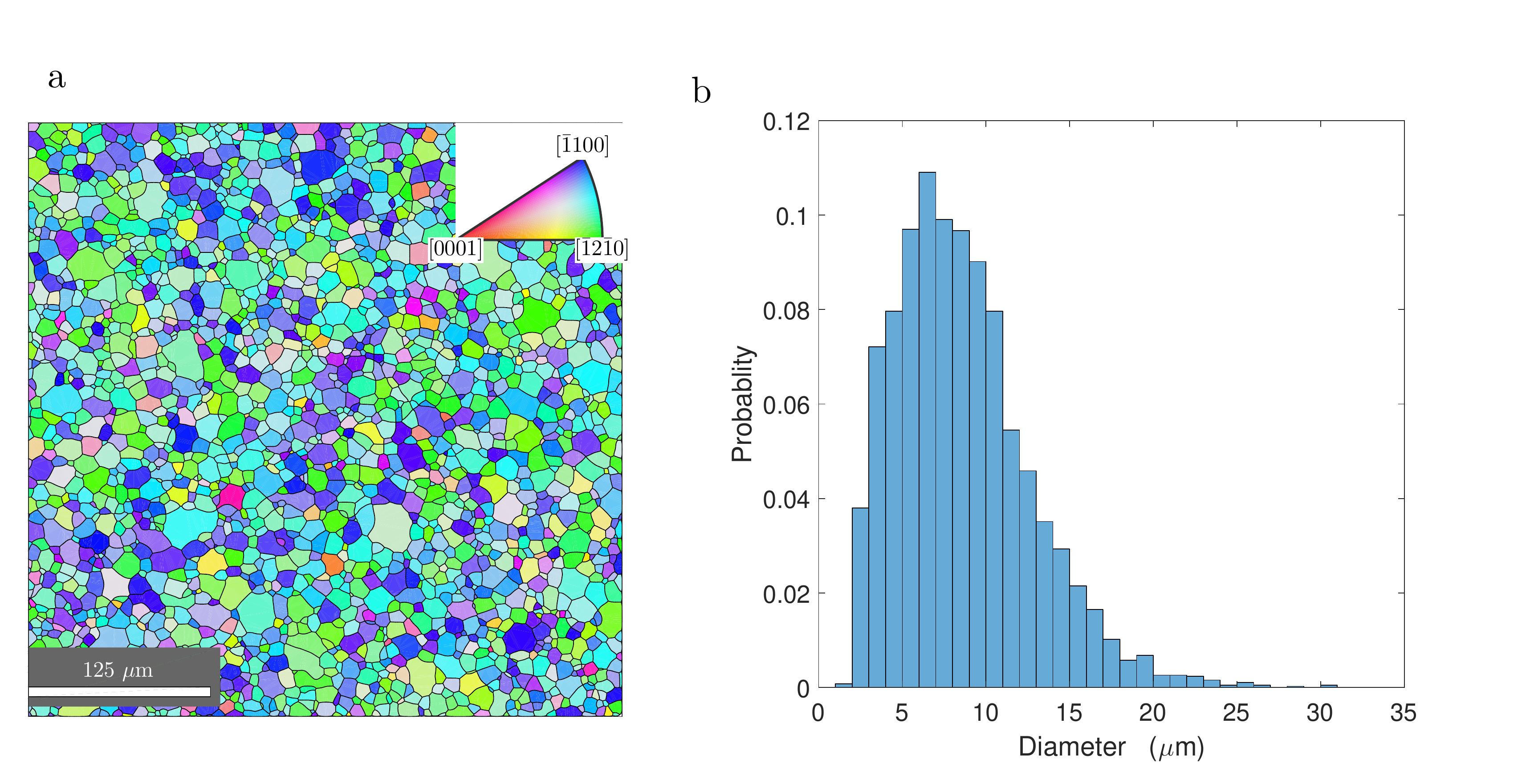}
\caption{(a) EBSD map of a cross-section perpendicular to the extrusion
direction. (b) Grain size distribution perpendicular to the extrusion
direction.}
\label{fig:EBSD_image}
\end{figure}

The microstructure of the material was analyzed by EBSD in a section
perpendicular to the extrusion direction and the corresponding inverse pole
figure map is shown in Fig. \ref{fig:EBSD_image}a. It shows a fully
recrystallized microstructure containing fine equiaxed grains. The grain
size distribution (determined from the area of each grain assuming that the
grain shape was circular) is shown in Fig. \ref{fig:EBSD_image}b and the
average grain size was 8.6 $\pm$ 4.0 $\mu$m.

\begin{figure}[tbp]
\centering
\includegraphics[width=0.9\textwidth]{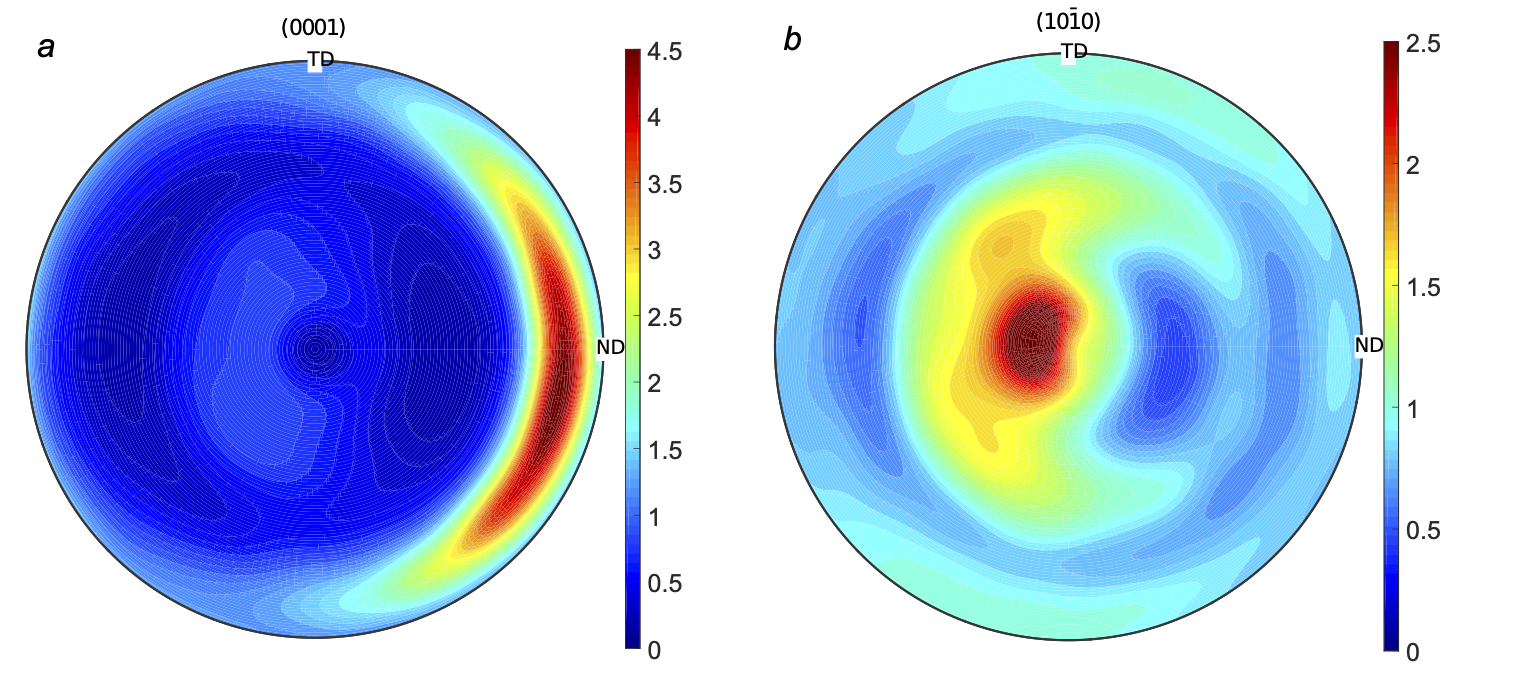}
\caption{Experimental pole figures indicating the orientation of basal
(0001) and (10$\bar{1}$0) planes. The numbers in the legend stand for
multiples of random distribution.}
\label{fig:poleFigureExp}
\end{figure}

The texture of the material obtained by X-ray diffraction is given by the
pole figures, indicating the orientation of basal (0001) and prismatic (10$\bar{1}$0])
planes (Fig. \ref{fig:poleFigureExp}). They show that the alloy has a
noticeable fiber texture component (very likely because the Nd content was
limited to 0.5 wt. \%. Assuming a criterion of a misorientation angle $\le $ 5$^{\circ}$, this microstructure can be approximated by a "matrix" of
Mg grains with random orientations (that occupies 87\% of the volume
fraction) which encompasses many "inclusions" with volume fraction of $13\%$
made up by Mg grains with the basal plane practically parallel to the
extrusion axis.

\begin{figure}[tbp]
\centering
\includegraphics[width=0.9\columnwidth]{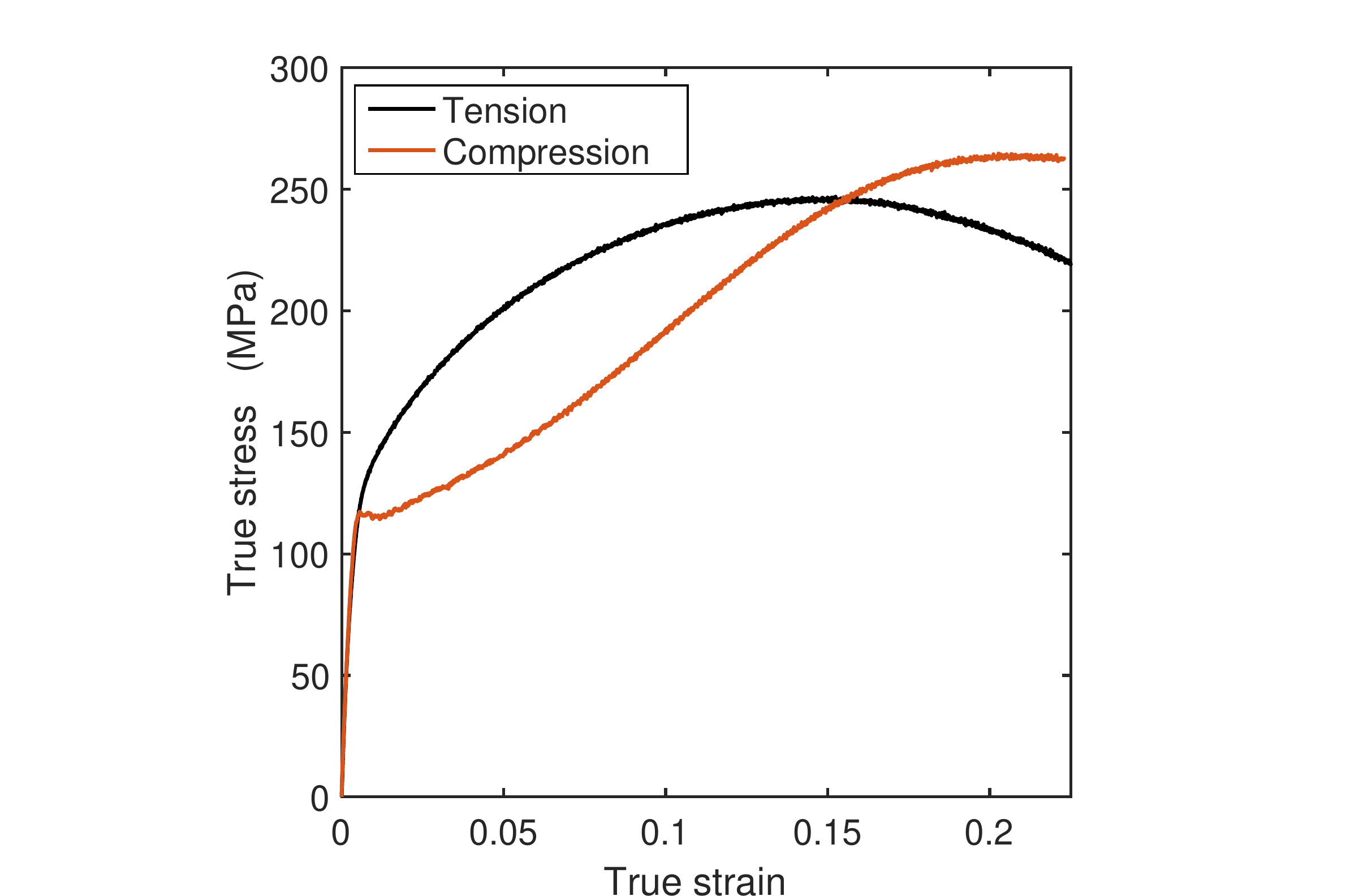}
\caption{Experimental tension and compression stress-strain curves along the
extrusion direction.}
\label{fig:EXP_monoSEcurvers}
\end{figure}

The mechanical properties of the alloy in tension and compression along the
extrusion axis are plotted in Fig. \ref{fig:EXP_monoSEcurvers}. They are
typical of a textured Mg alloy. The orientation of the basal planes parallel
to the extrusion axis facilitates twinning deformation in compression,
leading to the concave shape of the stress-strain curve after yielding.
However, extension twining is not favored by texture during tensile
deformation and deformation along the $c$ axis of the Mg lattice has to be
accommodated by pyramidal slip, leading to the parabolic hardening in the
stress-strain curves. Of course, basal slip will be activated both in
tension and compression (although the Schmid factor in most basal planes
will be low due to the texture) because of the low values of the critical
resolved shear stress for basal slip.

\begin{figure}[tbp]
\centering
\includegraphics[width=\textwidth]{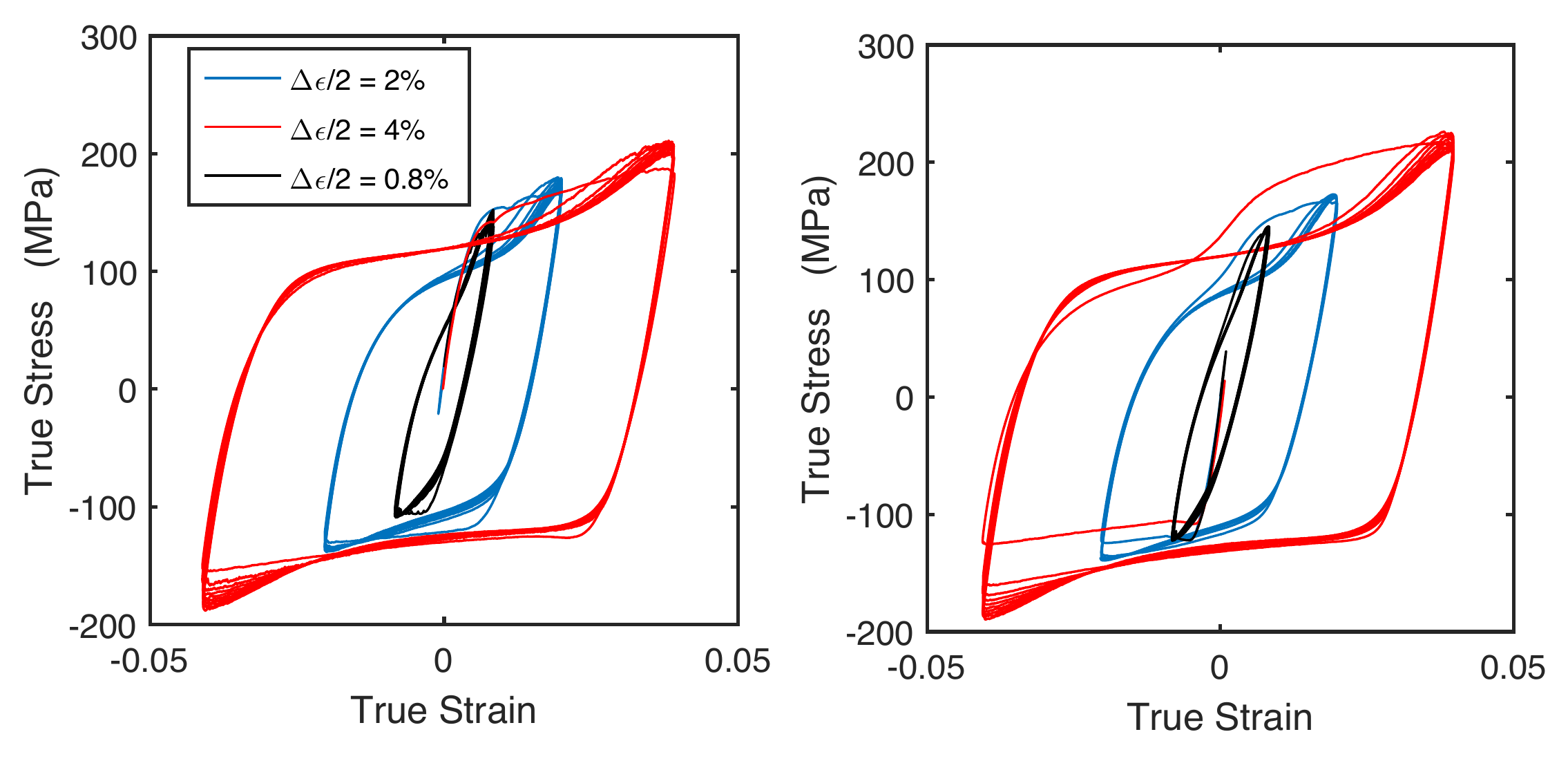}
\caption{Cyclic stress-strain curves of the specimens loaded un cyclic
deformation during the first ten cycles. (a) Specimens initially deformed in tension. (b) Specimens
initially deformed in compression.}
\label{fig:EXP_fatigueSEcurve}
\end{figure}

The cyclic stress-strain curves for different values of the applied cyclic
strain amplitude are plotted in Figs. \ref{fig:EXP_fatigueSEcurve}a and b.
The curves of the former figure correspond to fatigue tests in which the
specimen was initially deformed in tension and those in the latter to the
specimens initially deformed in compression. In general, the cyclic
stress-strain curves reached a steady-state condition after a few cycles and
presented similar features to those reported by \citep{WJB08, YZJ11,
Dong2014, Xiong2014} in Mg and Mg alloys. The specimens deformed initially
in tension (Fig. \ref{fig:EXP_fatigueSEcurve}a) showed parabolic hardening
which was followed by twinning when the load was reversed. The initial shape
of the stress-strain curve in the second load cycle in tension was not
parabolic shape but presented an S-shape because of the activation of
detwinning. Afterwords, the cyclic stress-strain curves did not change
significantly with the number of cycles until failure. Similar differences
in the stress-strain curves between the stress-strain curves of the first
and following cycles were found in the samples deformed initially in
compression (Fig. \ref{fig:EXP_fatigueSEcurve}b) although the sequence of
events was different: twinning developed during the initial deformation in
compression and detwinning appeared in the second part of the first cycle,
when the sample was loaded in tension.

\begin{figure}[tbp]
\centering
\includegraphics[width=\textwidth]{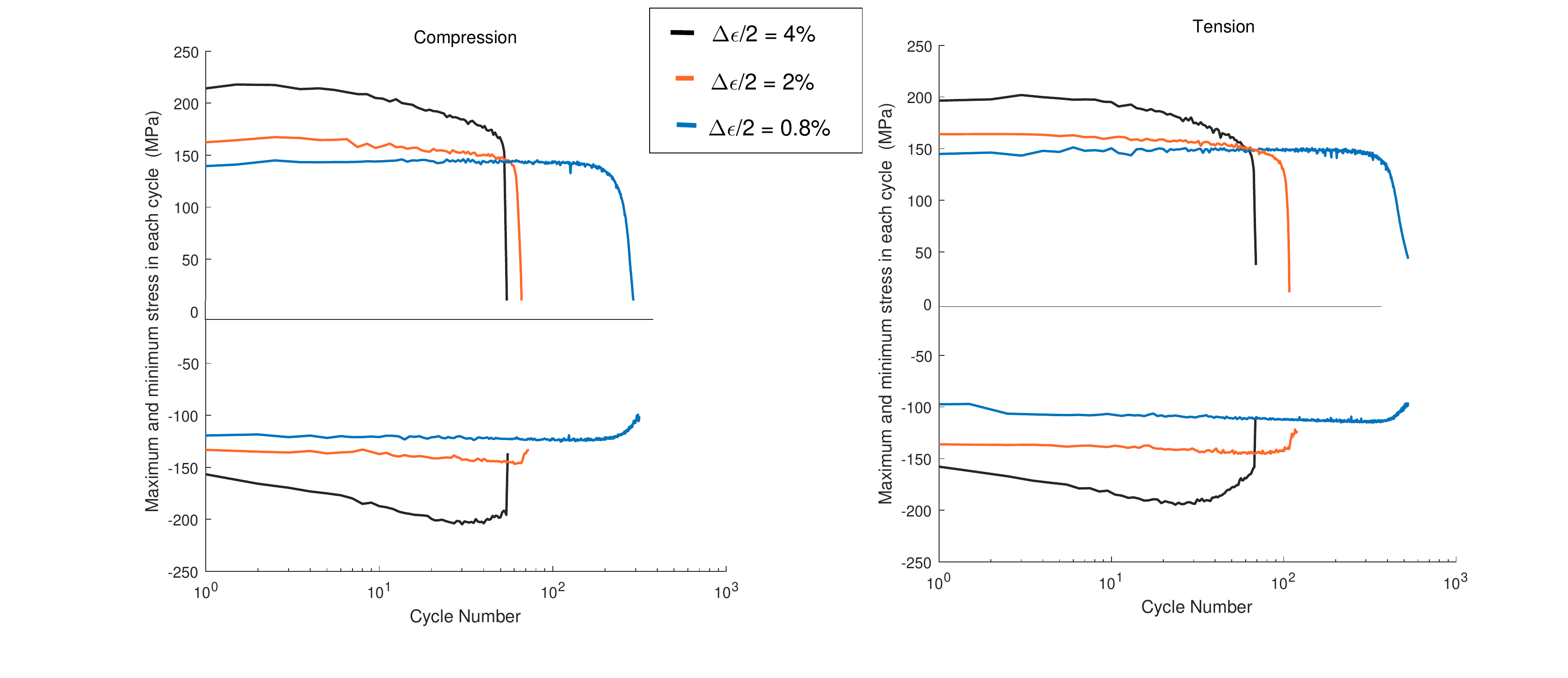}
\caption{Evolution of the maximum ($\sigma_{max}$) and minimum ($\sigma_{min}$) stress as a function of the number of cycles $N$.  (a) Specimens initially deformed in compression. (b) Specimens initially deformed in tension.}
\label{fig:SmaxSmin-N}
\end{figure}

The evolution of the maximum ($\sigma_{max}$) and minimum ($\sigma_{min}$)  stress with the number of fatigue cycles $N$ is plotted in Fig. \ref{fig:SmaxSmin-N}a) and b) for the specimens initially deformed in tension or in compression, respectively. $\sigma_{max} > |\sigma_{min} |$ for any given cyclic strain amplitude because twinning took place during the compressive part of the fatigue cycle. Moreover, the differences between the absolute values of $\sigma_{max}$ and $\sigma_{min}$ increased with $\Delta\epsilon$ because the contribution of twinning also increased. Both $\sigma_{max}$ and $\sigma_{min}$ remained practically constant throughout the test at $\Delta\epsilon/2$  = 0.8\%, while strain softening in tension and strain hardening in compression with the number of cycles was observed in the specimens deformed at $\Delta\epsilon/2$= 4\%. The specimens deformed at $\Delta\epsilon/2$= 2\% also showed slight softening in tension. The progressive hardening under compression can be rationalized by the presence of twins, which act as obstacles to dislocation motion, and increased with the cyclic strain amplitude because together with the volume fraction of twins. On the contrary, detwinning took place during the tensile part of the fatigue cycle, leading to a reduction in the stress necessary to move dislocations.

\begin{figure}[h!]
\centering
\includegraphics[width=0.6\textwidth]{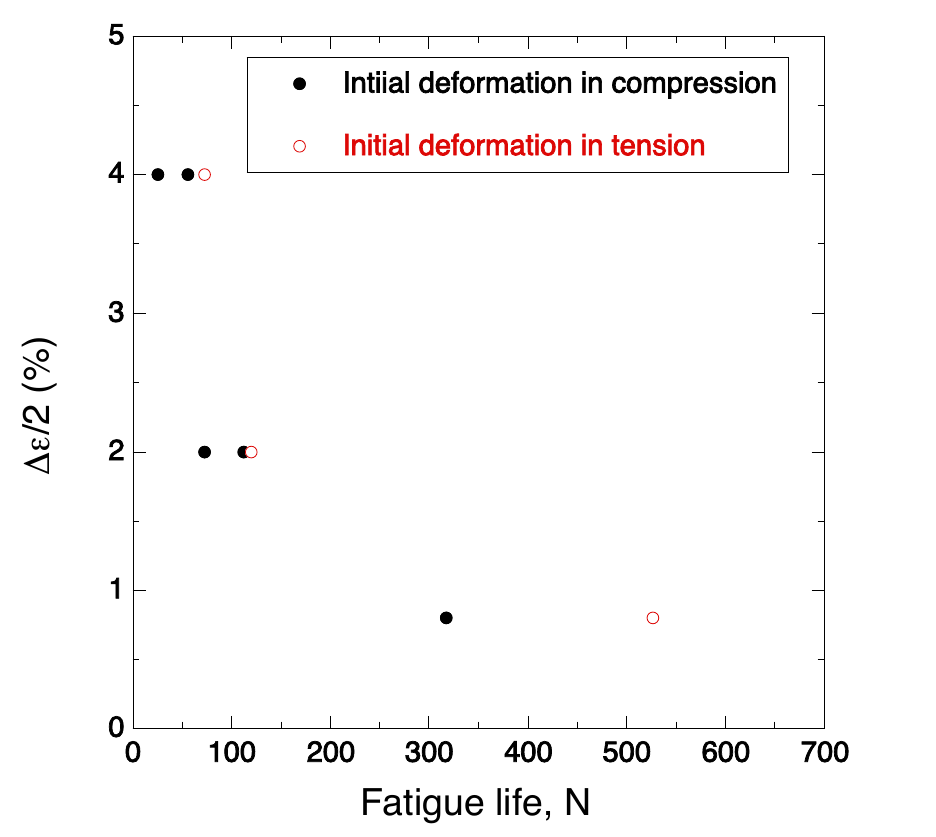}
\caption{Fatigue life of as a function of the cyclic strain semi-amplitude, $%
\Delta\protect\epsilon$/2 of the specimens initially deformed in tension and
in compression}
\label{fig:EXP_fatigueLife}
\end{figure}

The fatigue life of the tested specimens was plotted in Fig. \ref{fig:EXP_fatigueLife} as a function of the applied cyclic strain
semi-amplitude, $\Delta\epsilon/2$. It should be noted that the cyclic stress-strain curves in Figs. \ref{fig:EXP_fatigueSEcurve} and \ref{fig:SmaxSmin-N} were very similar for both tests.

\begin{figure}
\centering
\includegraphics[width=1.0\textwidth]{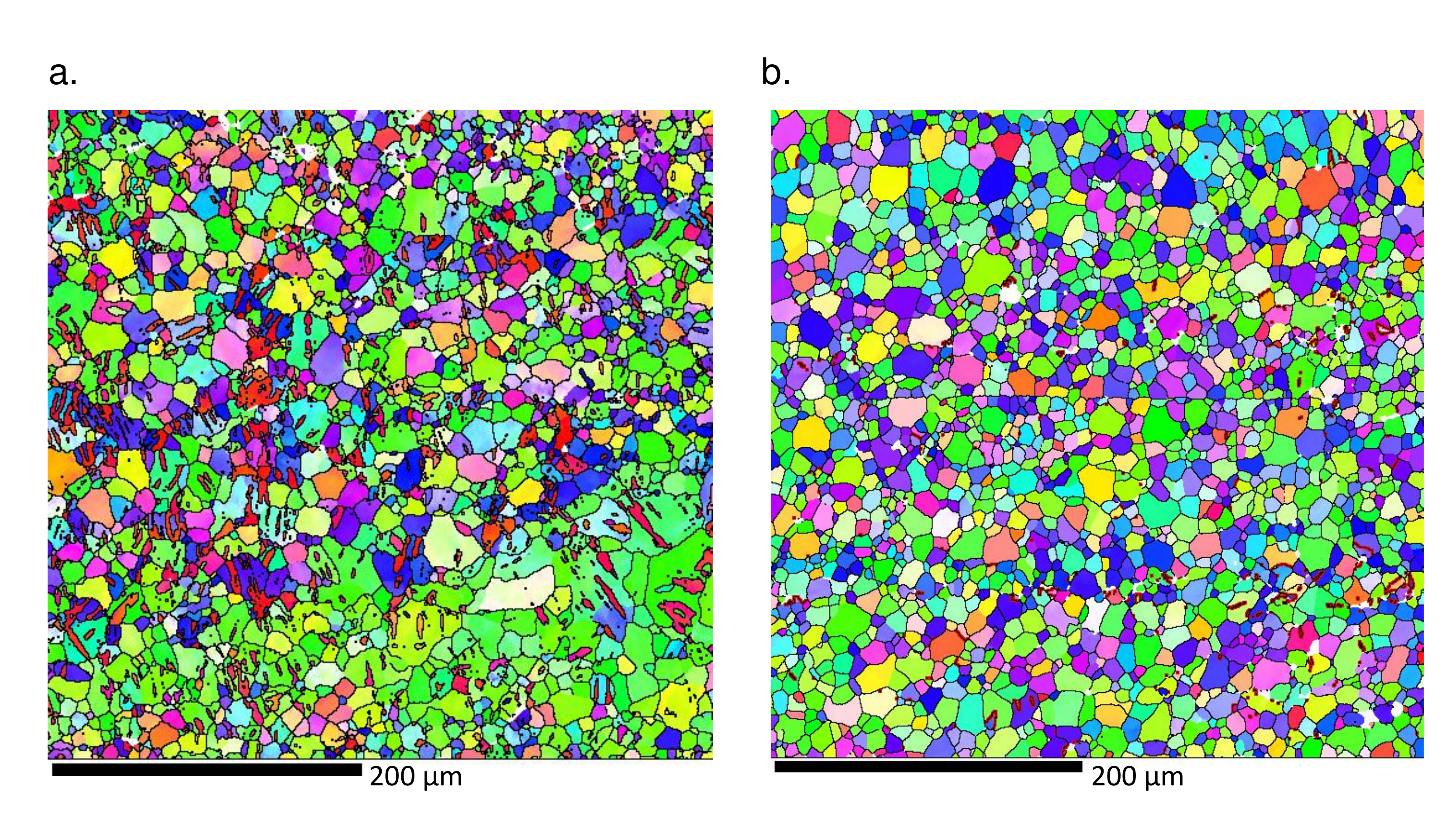}
\caption{EBSD map of the longitudinal section of the specimen deformed at $\Delta\epsilon $ = 2\% (beginning in tension). (a) Close to the fracture surface. (b) Far away from the fracture surface. 
The extrusion direction is horizontal. The twinned regions appear as red zones in the EBSD map.} 
\label{fig:twin_T02}
\end{figure}

Selected specimens were sliced parallel to the loading direction after fracture and the microstructure was analyzed with EBSD close and far-away from the fracture surface. The EBSD images were analyzed using the HKL Channel 5 software and the $(10\bar{1}2)$ tension twin regions were determined from the misorientation angle of $86^{\circ}$ between the Mg matrix and the twinned regions (Fig. \ref{fig:twin_T02}). The twin volume fraction was determined close and far-away from the fracture surfaces and is depicted in Table \ref{tab:TVF}. It should be noted that the fraction of twinned material is always higher near to the fracture surface indicating that the localization of damage in this region influenced the development of twins. In addition, twinning took place during the compressive part of the cycle while detwinning occurred in tension. Thus,  the twin area fraction was much larger in the specimens that failed in compression as compared with those that failed in tension for $\Delta\epsilon/2=2\%$. Unfortunately, all the specimens deformed at $\Delta\epsilon/2=4\%$ failed in compression and those tested at $\Delta\epsilon/2=0.8\%$ failed in tension. 

\begin{table}[tbp]
\centering
\begin{tabular}{|c|c|c|c|c|}
\hline
$\Delta\epsilon/2$ & initial  & TAF near fracture  & TAF center   & failure  \\ 
(\%)  &  loading  & (\%) & (\%) &  point   \\ \hline
4\%   & Compression & not measured  & 36.45 & compression \\ 
4\%   & Tension & not measured  & 26.56 & compression \\ 
2\%   & Compression  & 35.5 & 19.6 & compression \\ 
2\%   & Tension & 9.8 & 0.3  & tension \\ 
0.8\% & Compression  & 5.9 & 0.2  & tension \\ 
0.8\% & Tension  & 8.8 & 1.5  & tension \\ \hline
\end{tabular}%
\caption{Twin area fraction (TAF) near and far-way from the fracture surfaces measured from EBSD maps on longitudinal sections of the samples deformed with different cyclic strain semi-amplitudes $\Delta\epsilon/2$. The initial loading direction (either tension or compression) and whether the specimen failed close to the maximum tensile strain or to the minimum compressive strain is indicated for each test.}
\label{tab:TVF}
\end{table}

\section{Computational homogenization}

In literature the cyclic stress strain curves as well as twin volume fraction evolutions of Mg alloys were widely
simulated by crystal plasticity models based on finite element, fast Fourier transformation and self consistent approaches \citep{HWang2013, Segurado2018, Paramatmuni2019, Tang2019, Indurkar2020}. Most of these CP models adopted phenomenological type flow and hardening laws where twinning was modeled by as a pseudo-slip deformation mode \citep{Kalidindi1998, Herrera-Solaz2014a,
Herrera-Solaz2014b}.

\subsection{Crystal plasticity framework}

The crystal plasticity model followed the standard multiplicative
decomposition of the deformation gradient $\mathbf{F}$ into the elastic ($
\mathbf{F}_{e}$) and plastic ($\mathbf{F}_{p}$) components according to 
\citep{Kalidindi1992}

\begin{equation}
\mathbf{F}=\mathbf{F}_{e}\mathbf{F}_{p}.
\end{equation}

In the intermediate configuration, the plastic velocity gradient $\mathbf{L}%
_{\text{p}}$ is the sum of the contributions due to slip ($\mathbf{L}%
_{p}^{sl}$) and twinning ($\mathbf{L}_{p}^{tw}$) as 

\begin{equation}
\mathbf{L}_{p}=\mathbf{L}_{p}^{sl}+\mathbf{L}_{p}^{tw}
\end{equation}

\noindent and plastic slip in the twinned regions of the crystal was not
included in the model because it is unlikely to occur due to the continuous
twinning-detwinning during cyclic deformation \citep{Aeriel2019-twdtw}.

Plastic deformation in Mg alloys is known to occur in 3 $<a>$ (0001) [11$%
\bar 2$0] basal, 3 $<a>$ (1$\bar 1$00) [11$\bar 2$0] prismatic and 12 $<a+c>$
(10$\bar 1$1) [11$\bar 2$3] pyramidal slip systems, while tensile twinning
takes place along 6 (01$\bar 1$2) [0$\bar 1$11] twins variants. The
contribution of plastic slip to the plastic velocity gradient is expressed
as 
\begin{equation}
\mathbf{L}_{p}^{sl}=\left( 1-\overset{N_{tw}}{\underset{\beta =1}{\sum }}%
f_{\beta }\right) \overset{N_{sl}}{\underset{\alpha =1}{\sum }}\dot{\gamma}%
_{\alpha }\hat{\mathbf{s}}_{\alpha }^{sl}\otimes \hat{\mathbf{m}}_{\alpha
}^{sl}  \label{velo_gradsl}
\end{equation}

\noindent where $N_{sl}$ and $N_{tw}$ stand for the number of slip and twin
systems, respectively. $\hat{\mathbf{s}}_{\alpha }^{sl}$ and $\hat{\mathbf{m}%
}_{\alpha }^{sl}$ are unit vectors parallel to the slip direction and slip
plane normal, respectively, of slip system $\alpha $ in the reference
configuration and $f_{\beta }$ is the volume fraction of twins corresponding
to the twin system $\beta $.

The contribution of twining to the plastic velocity gradient can be
expressed as

\begin{equation}
\mathbf{L}_{p}^{tw}= \overset{N_{tw}}{\underset{\beta=1}{\sum}}\dot{f}%
_{\beta}\gamma_{tw} \hat{\mathbf{s}}_{\beta}^{tw}\otimes \hat{\mathbf{m}}%
_{\beta }^{tw}.  \label{velo_grad_tw}
\end{equation}

\noindent where $\hat{\mathbf{s}}_{\beta}^{tw}$ and $\hat{\mathbf{m}}_{\beta
}^{tw}$ are unit vectors parallel to the twin direction and twin plane
normal, respectively, of twin system $\beta$ in the reference configuration
and $\gamma_{tw}$ (= 0.129 in Mg) \citep{Herrera-Solaz2014a} is the
eigenstrain associated to extension twinning which only takes place when the deformation
leads to an extension of the $c$ axis of the HCP lattice.

The slip rate in a slip system $\alpha$ is defined by a power-law dependency
according to

\begin{equation}
\dot{\gamma}_{\alpha}=\dot{\gamma}_{0} \left[ \frac{\tau_{\alpha}-\tau_{
\alpha}^{b}}{g_{\alpha}} \right]^{\frac{1}{m}} \mathrm{sign}
(\tau_{\alpha}-\tau_{\alpha}^{b})  \label{dgmdt_slip}
\end{equation}

\noindent where $\dot{\gamma}_{0}$ is the reference shear rate parameter, $m$
the strain rate sensitivity exponent and $g_{\alpha}$ and $\tau_{\alpha}^{b}$ stand
for the isotropic and kinematic hardening contributions, respectively. $%
\tau_\alpha$ is the resolved shear stress on the slip system $\alpha$ that
can be expressed as the projection of the second Piola-Kirchhoff stress, $\mathbf{S}
$, on the current slip system $\alpha$, which is given by

\begin{equation}  \label{eq:rss}
\tau_\alpha=\mathbf{S}:(\hat{\mathbf{s}}_{\alpha}^{sl}\otimes \hat{\mathbf{m}%
} _{\alpha}^{sl}).
\end{equation}

The evolution law of the isotropic hardening can be expressed as \citep{Kalidindi1998}

\begin{equation}
\dot{g}_{\alpha }=\underset{\gamma =1}{\overset{N_{\text{sl}}}{\sum }}
h_{\alpha \gamma }^{\prime }H_{\gamma }\bigg(1-\frac{\tau _{\gamma}}{
g_{\gamma }^{sat}}\bigg)^{a_{ss}}|\dot{\gamma}_{\gamma }|+\underset{\beta =1}
{\overset{N_{\text{tw}}}{\sum }}h_{\alpha \beta }^{\prime \prime }H_{\beta }
\bigg(1-\frac{\tau _{\beta }}{g_{\beta }^{sat}}\bigg)^{a_{st}}|\dot{\gamma}
_{\beta }|  \label{dtaucdt_slip}
\end{equation}

\noindent where $h_{\alpha \gamma}^{\prime}$ is the latent hardening
parameter between slip system $\alpha$ and slip system $\gamma $. Parameter $h_{\alpha \beta
}^{\prime \prime}$ stands for the latent hardening parameter between slip
system $\alpha$ and twin system $\beta$. Parameters $g_{\gamma}^{sat}$ and $H_{\gamma}$
stand for the saturation critical resolved shear stress and the hardening
modulus, respectively, of the slip system $\gamma $ while $g_{\beta }^{sat}$
and $H_{\beta}$ are the corresponding values for the twin system $\beta $. $
a_{ss}$ is the slip-slip hardening exponent, while $a_{st}$ is the twin-slip
hardening exponent.

The evolution of the backstress $\tau_{\alpha}^{b}$ which determines the
kinematic hardening during cyclic deformation is a simplification of the
Ohno-Wang macroscopic model \citep{Ohno1993}, which is able to reproduce the
complex cyclic behavior of a polycrystal. Mathematically,

\begin{equation}
\dot{\tau}_{\alpha }^{b}=c_{\alpha }\dot{\gamma}_{\alpha }-d_{\alpha }\tau
_{\alpha }^{b}\left( \frac{|\tau _{\alpha }^{b}|}{c_{\alpha }/d_{\alpha }}%
\right) ^{k_{\alpha }}|\dot{\gamma}_{\alpha }|  \label{backstress_rate}
\end{equation}

\noindent where the parameter $c_{\alpha }/d_{\alpha }$ indicates the
stabilized scalar back-stress for slip system $\alpha $, $1/d_{\alpha }$
stands for the absolute value of the shear $|\gamma _{\alpha }|$ when the
back-stress stabilizes and $k_{\alpha }$ accounts for the non-linearity of
the back stress evolution. This model was successfully applied to simulate
the evolution of kinematic hardening in Ni-based superalloys during
computational homogenization of polycrystals \citep{Cruzado2017,Cruzado2018}.

As twinning systems were treated as pseudo-slip systems, the evolution of
the twin volume fraction in each twin variant, $\dot{f}_{\beta}$, also follows
a power-law but the peculiarities of twinning have to be taken into account. In
particular, tension twinning can only occur when the resolved shear stress
on the twin plane leads to an extension of the $c$ axis and
cannot progress anymore in any twin system if the whole region associated to
the Gauss point is completely twinned (i.e. $\sum_{\beta} f_{\beta}$ = 1). In
addition, the twinned region can detwin when the resolved shear stress is
opposed to the one required for twinning and reaches a critical value %
\citep{ZJS19, MPB19, Briffod2019, Yaghoobi2020}. This means that $\dot{f}_{\beta}$ can be
negative or positive, but the resistance for twin boundary motion under both
scenarios can be different. Mathematically, these conditions can be taken
into account as

\begin{equation}
\left\{ 
\begin{array}{ll}
\dot{f}_{\beta }=\dot{f}_{0}\left( \frac{|\tau _{\beta }|}{g_{\beta }}%
\right) ^{\frac{1}{m}}sign(\tau _{\beta })\quad & \mathrm{if}\quad f_{\beta
}>0\mathrm{\ }\text{and}\ \ \sum f_{\beta }<1 \\ 
\dot{f}_{\beta }=0\quad & \mathrm{if}\quad \tau _{\beta }\geqslant 0\text{\
and \ }\sum f_{\beta }\approx 1 \\ 
\dot{f}_{\beta }=0\quad & \mathrm{if}\quad \tau _{\beta }<0\text{\ and \ }%
f_{\beta }\approx 0 \\ 
& 
\end{array}
\right.  \label{dgmdt_twin}
\end{equation}

\noindent where $\dot{f}_{0}$ is the reference twin volume fraction rate, $%
\tau _{\beta }$ represents the resolved shear stress on twin system, which
can be calculated as in eq. \eqref{eq:rss} for the corresponding twin
system, and $g_{\beta}$ are the stresses necessary to activate twining or
detwinning. They do not have to be equal, following recent experimental
observations \citep{ZJS19, MPB19}.

The hardening law for twin boundary motion is expressed as

\begin{equation}
\dot{g}_{\beta }=\underset{\beta =1}{\overset{N_{\text{tw}}}{\sum }}%
h_{\alpha \beta }^{\prime \prime \prime }H_{\beta }\gamma _{tw}\left( 1-%
\frac{\tau _{\beta }}{g_{\beta }^{sat}}\right) ^{a_{tt}}|\dot{f}_{\beta }|
\label{dtaucdt_twin}
\end{equation}

\noindent where -- similar to equation (\ref{dtaucdt_slip}) --- $h_{\alpha
\beta }^{\prime \prime \prime }$ stands for the twin/twin latent hardening
parameter, $H_{\beta }$ the corresponding hardening modulus, $g_{\beta
}^{sat}$ the saturated critical resolved shear stress for twinning or
de-twinning, and $a_{tt}$ the twin-to-twin hardening exponent. It is
assumed that plastic slip does not influence the hardening of twinning.

\subsection{Homogenization strategy}

\begin{figure}[tbp]
\centering
\includegraphics[width=0.9\textwidth]{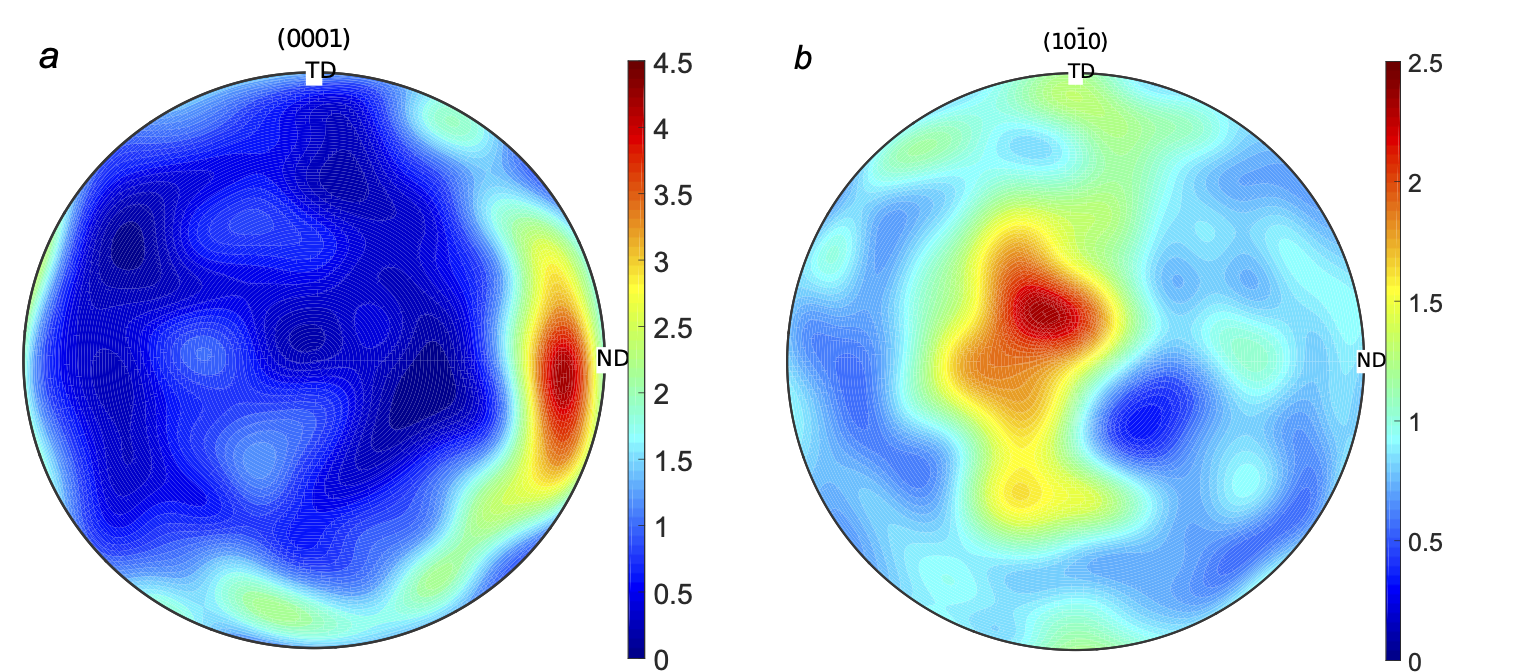}
\caption{Pole figures of the RVE of the polycrystal indicating the
orientation of basal (0001) and (10$\bar{1}$0) planes. The numbers in the
legend stand for multiples of random distribution.}
\label{fig:poleFigureSim}
\end{figure}

The mechanical response of the polycrystalline Mg alloys was determined by
means of the finite element simulation of a Representative Volume Element
(RVE) of the microstructure. The RVE was created using NEPER \citep{Neper} included 200 grains which followed
the grain size distribution in Figure \ref{fig:EBSD_image}b). The polycrystal was represented by a cubic
domain, which was discretized with a regular mesh of 11 x 11 x 11 cubic C3D8R
finite elements where each crystal is represented by about 7 elements. The
orientation of the grains was obtained from the experimental orientation
distribution function obtained by X-ray diffraction (Fig. \ref
{fig:poleFigureExp}). The actual pole figures of the grains in the RVE
indicating the orientation of basal (0001) and (10$\bar{1}$0) planes are
shown in Fig. \ref{fig:poleFigureSim}. This RVE was used in section 5.1 to identify the parameters of the crystal plasticity model and in section 5.2 to identify the cyclic deformation mechanisms.

The mechanical response of the aforementioned RVE under uniaxial cyclic deformation
was simulated using Abaqus/Standard \citep{A20} within the framework of the
finite deformations theory with the initial unstressed state as reference.
The behavior of each crystal followed the crystal plasticity model presented above that was
implemented as User Material subroutine (UMAT) in Abaqus. Periodic boundary
conditions were applied along the three directions of the RVE. Uniaxial
cyclic deformation was obtained by applying an alternating cyclic
displacement in one direction while the total stresses in both perpendicular
directions were 0. The applied strain rate was 10$^{-3}$ s$^{-1}$. More
details about the numerical homogenization strategy can be found elsewhere 
\citep{HSL15}

\section{Numerical results and discussion}

\subsection{Parameter identification}

The elastic constants of the Mg single crystals for the simulations were
obtained from the literature \citep{Zhang2012} and can be found in Table \ref
{elastic_constants}. The main parameters of the crystal plasticity model were adjusted by
comparison of experimental and simulated cyclic stress-strain curves. 
They were the initial and saturated critical resolved shear stress of each
slip system as well as of twinning and de-twinning, the corresponding strain
hardening moduli and the parameters for kinematic hardening for each slip
system. They are depicted in Table \ref{CP_paramter_part2}. Some model parameters were adopted from the literature \citep{AZ31_Teresa, Herrera-Solaz2014b}. They are included in Table \ref{CP_paramter_part1}.

\begin{table}[t]
\centering%
\begin{tabular}{c|c|c|c|c}
\hline
$C_{11}$ & $C_{12}$ & $C_{33}$ & $C_{13}$ & $C_{44}$ \\ \hline
59.4 & 25.6 & 61.6 & 21.4 & 16.4 \  \\ \hline
\end{tabular}%
\caption{Elastic constants (in GPa) after \cite{Zhang2012}.}
\label{elastic_constants}
\end{table}

\begin{table}[t!]
\centering%
\begin{tabular}{l|c|c|c|c|c}
\hline
Slip/twin mode & basal & prismatic & pyramidal & twinning & detwinning \\ \hline
$g^{ini}$ (MPa) & 8 & 78 & 165 & 30 & 23.7 \\ \hline
$g^{sat}$ (MPa) & 19 & 90 & 180 & 31 & 24.5 \\ \hline
$H$ (MPa) & 25 & 150 & 450 & 600 & 600 \\ \hline
$c$ (MPa) & 40 & 390 & 825 &  &  \\ \hline
$d$ & 5 & 5 & 5 &  &  \\ \hline
$k$ & 10 & 10 & 10 &  &  \\ \hline
\end{tabular}%
\caption{Model parameters determined by comparison of the simulation results
with the experimental cyclic stress-strain curves at different cyclic strain
semi-amplitudes.}
\label{CP_paramter_part2}
\end{table}

\begin{table}[t]
\centering%
\begin{tabular}{|l|l|}
\hline
$\dot{f}_{0}$ & 0.001 s$^{-1}$   \\ 
$\dot{\gamma}_{0}$ & 0.001 s$^{-1}$  \\ 
$m$ & 0.1   \\ 
$h_{\alpha \beta}^{\prime}$ (coplanar slip systems) & 1   \\ 
$h_{\alpha \beta}^{\prime}$ (non coplanar slip systems) & 1.4   \\ 
$h_{\alpha \beta}^{\prime\prime} $ & 2.0   \\ 
$h_{\alpha \beta}^{\prime\prime\prime} $ & 1.0   \\ 
$a_{\text{ss}} $ & 1.1   \\ 
$a_{\text{st}} $ & 2.0   \\ 
$a_{\text{tt}} $ & 1.0   \\ \hline
\end{tabular}%
\caption{Model parameters of flow and hardening laws for Mg alloys after \citep
{AZ31_Teresa, Herrera-Solaz2014b}.}
\label{CP_paramter_part1}
\end{table}

\begin{figure}[tbp]
\centering%
\includegraphics[width=0.9\textwidth]{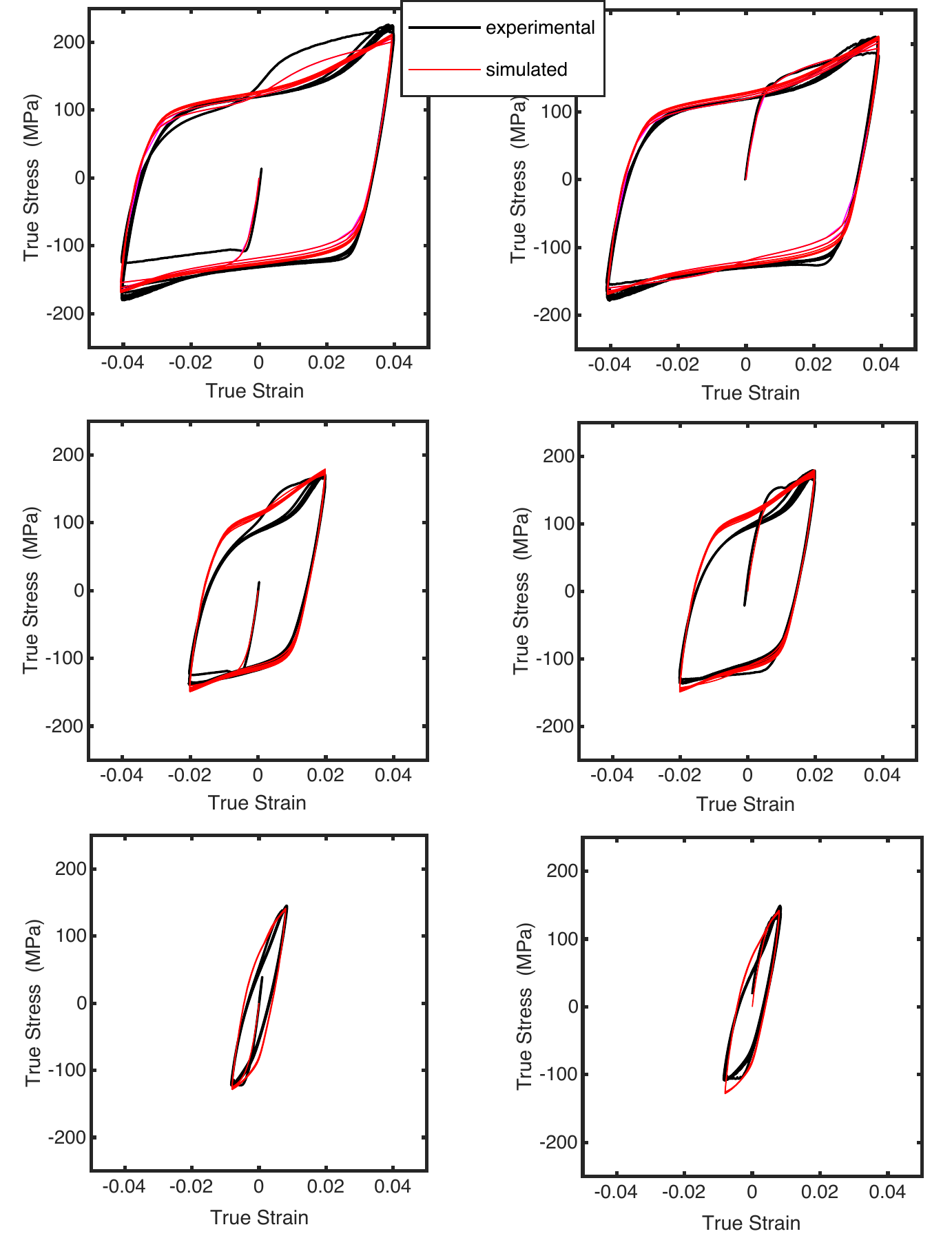}
\caption{Experimental (black) and simulated (red) cyclic stress-strain
curves of the Mg - 1Mn - 0.5Nd (wt.\%) alloy deformed along the extrusion
direction. Different plots correspond to specimens deformed initially in
either tension or compression at different cyclic strain semi-amplitudes ($
\Delta\protect\epsilon/2$ = 0.8\%, 2\% and 4\%).}
\label{fig:CPFEM_fitting}
\end{figure}

The experimental cyclic stress-strain curves are plotted together with those
obtained by computational homogenization in Fig. \ref{fig:CPFEM_fitting} for
the specimens deformed initially in either tension or compression at
different cyclic strain semi-amplitudes. Overall, the simulated cyclic
stress-strain curves are in good agreement with the experimental results and
capture the anisotropy in the curves associated with the successive
development of twinning and detwinning during cyclic loading. It is worth
noting that the experimental and predicted values of the saturated maximum
(tension) and minimum (compression) stresses were similar in the case of the
samples deformed initially in tension or in compression for each cyclic
strain amplitude. Moreover, the stabilization of the simulated cyclic stress-strain curves occurred by rapidly (in a few cycles) while there was some variation during the fatigue life in the experimental results, particularly in the specimens deformed at $\Delta\epsilon/2$ = 4\% (Fig. \ref{fig:SmaxSmin-N}).

In addition to the stress-strain curves, the fraction of twinned material predicted by the computational homogenization strategy is compared in Table \ref{tab:TVF-ES} with the experimental measurements of the twin area fraction far away from the fracture surfaces. It is important to notice that the fraction of twinned material was much larger when the sample failed in compression than in tension in the sample deformed at $\Delta\epsilon$ = 2\% because twinning developed during the compressive part of the fatigue cycle and detwinning in the tensile one. The numerical model was able to capture accurately this phenomenon particular for the specimens deformed at  $\Delta\epsilon/2$=2\%, as shown by the results in Table \ref{tab:TVF-ES}. Unfortunately, the same result could not be ascertained in the samples deformed at $\Delta\epsilon$ = 4\% and 0.8\% because the former always failed in compression and the latter in tension.

\begin{table}[tbp]
\centering
\begin{tabular}{|c|c|c|c|c|}
\hline
$\Delta\epsilon/2$   &  Experimental   & Simulated  & Failure point\\ 
(\%)  &  (\%) & (\%) &    \\ \hline
4\%   &  36.45 & 30.2 & compression \\ 
4\%   &  26.56 & 30.1 & compression \\ 
2\%   &  19.6 & 19.13 & compression \\ 
2\%   &  0.3 &  7.2 & tension \\ 
0.8\% &  0.2 &  2.4 & tension \\ 
0.8\% &  1.5 &  2.5 & tension \\ \hline
\end{tabular}%
\caption{Experimental and simulated twin fraction in samples deformed at different cyclic strain semi-amplitudes $\Delta\epsilon/2$. The failure point (whether the specimen failed close to the maximum tensile strain or to the minimum compressive strain) is indicated for each test.}
\label{tab:TVF-ES}
\end{table}

\subsection{Deformation mechanisms}

\begin{figure}[tbp]
\centering
\includegraphics[width=1.0\textwidth]{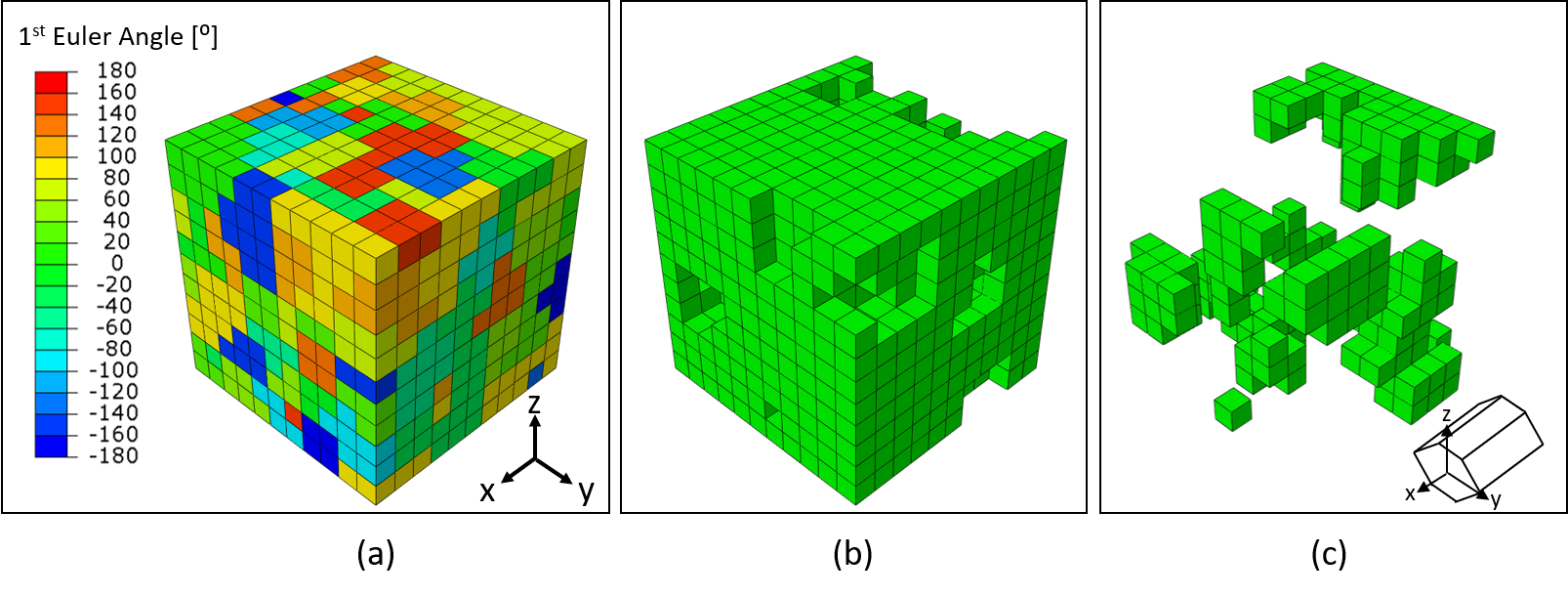}
\caption{RVE of the  extruded Mg-1Mn-0.5Nd alloy. (a) Magnitude of the first Euler angle 
in each grain of the RVE. (b) Grains with random
orientations (87\%) and (c) inclusions (13\%) oriented with the basal plane practically parallel to the extrusion and loading axis $Z$, as shown in the hcp unit cell.}
\label{fig:CPFEM_RVE}
\end{figure}

The extruded Mg-1Mn-0.5Nd alloy presented a peculiar texture (Fig. \ref{fig:poleFigureExp}) which could be approximated by a composite material formed by Mg inclusions with the basal plane practically parallel to the extrusion axis ($<$ 5$^\circ$) embedded in a matrix of Mg crystals with random texture. This texture was assigned to the the aforementioned RVE shown in Fig. \ref{fig:CPFEM_RVE} and it is expected that the deformation mechanisms in the matrix and inclusion regions of the RVE should be different. 

\begin{figure}[tbp]
\centering
\includegraphics[width = 0.48\textwidth]{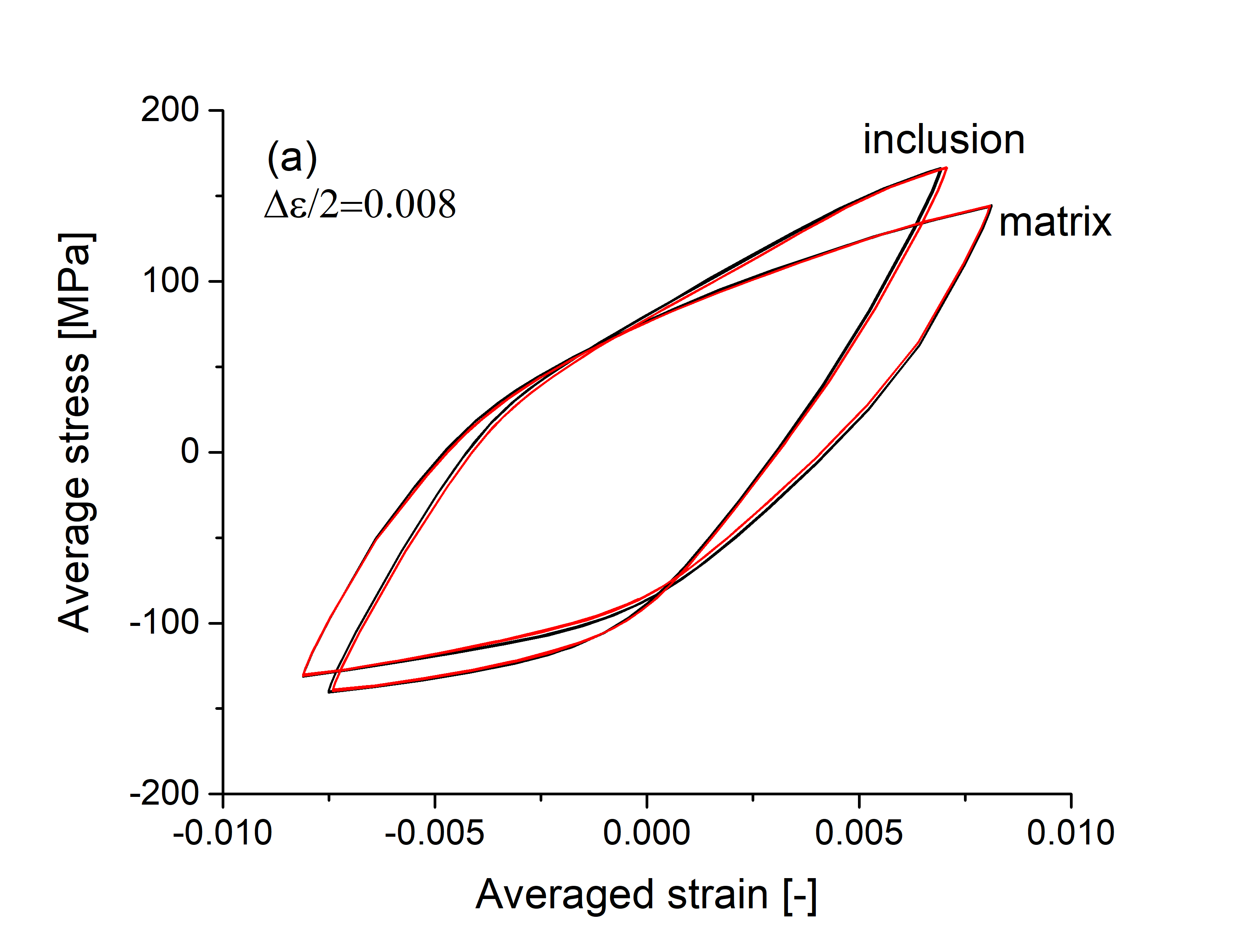}
\includegraphics[width = 0.48\textwidth]{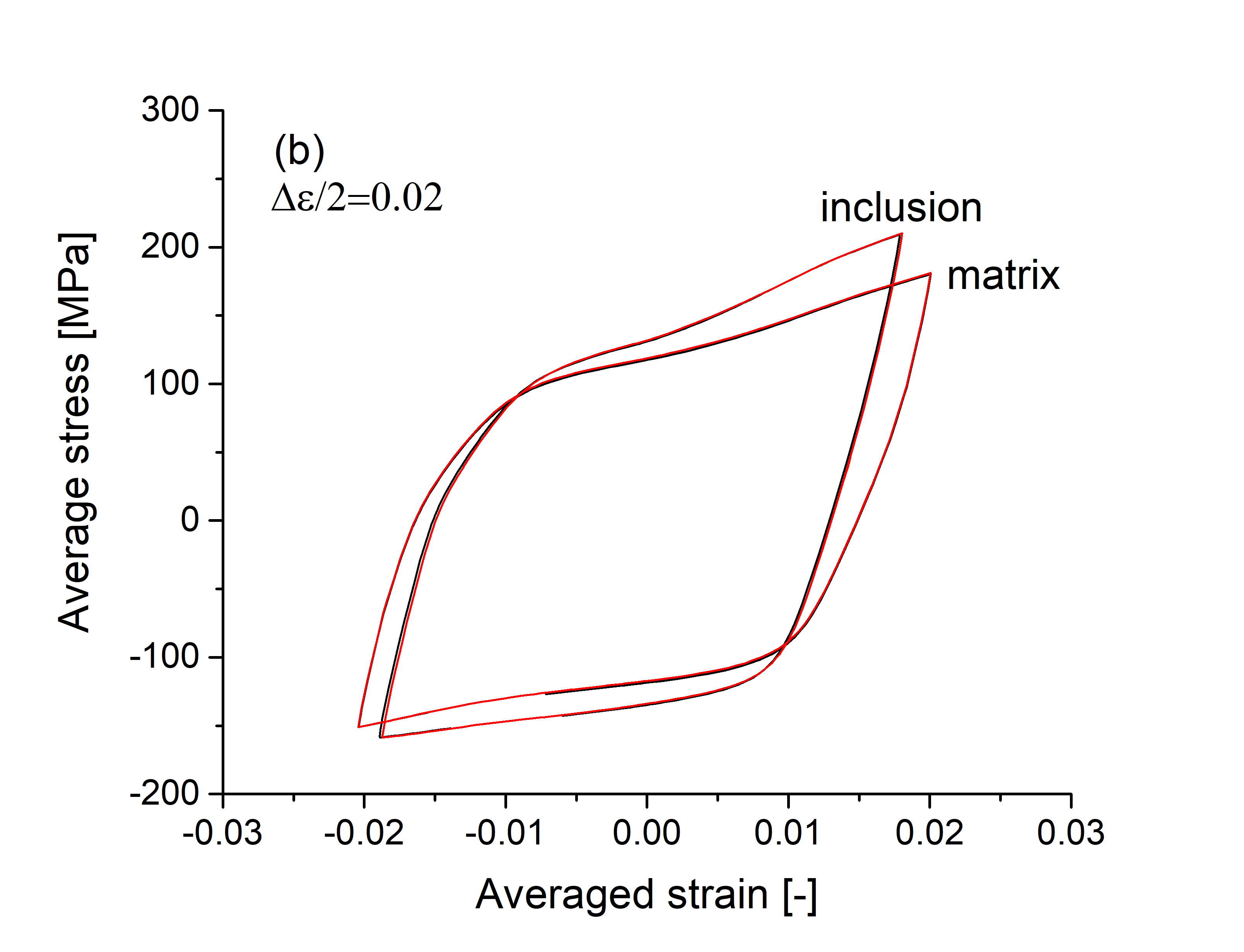} 
 \includegraphics[width = 0.48\textwidth]{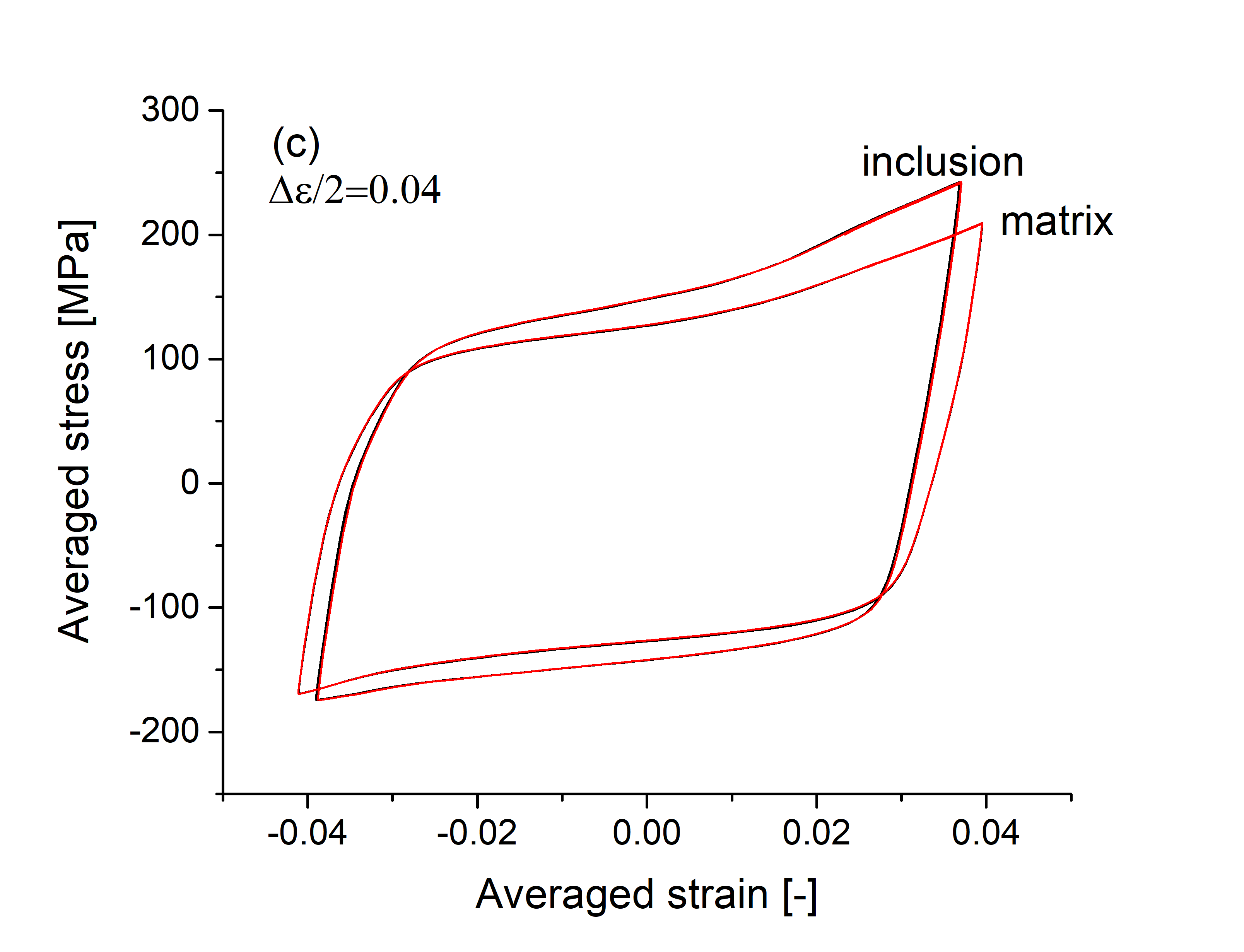} 
\caption{Average stable cyclic stress-strain curves in the matrix and inclusion regions of the RVEs for different values of the applied cyclic strain semi-amplitude. (a) $\Delta\protect\epsilon/2$ = 0.8\%. (b) $\Delta\protect\epsilon/2$ = 2.0\%. (c) $\Delta\protect\epsilon/2$ = 4.0\%. Curves obtained when the RVE was initially deformed in tension (red) or in compression (black) are presented.}
\label{fig:localSE}
\end{figure}

The average cyclic stress-strain curves corresponding to the matrix and the inclusions obtained by the simulation of the RVE are plotted in Figs.  \ref{fig:localSE}a, b and c for the specimens deformed at $\Delta\protect\epsilon/2$ = 0.8\%, 2\% and 4\%. The curves correspond to the saturated cyclic-strain curves and include the curves when the RVE was initially deformed in tension or compression, which were superposed in all cases. These curves show that the strain hardening rate in the inclusions during tensile part of the fatigue cycle was higher than that in the matrix while similar hardening rates were found in both regions in compression.

\begin{figure}[tbp]
\centering
\includegraphics[width = 1.0\textwidth]{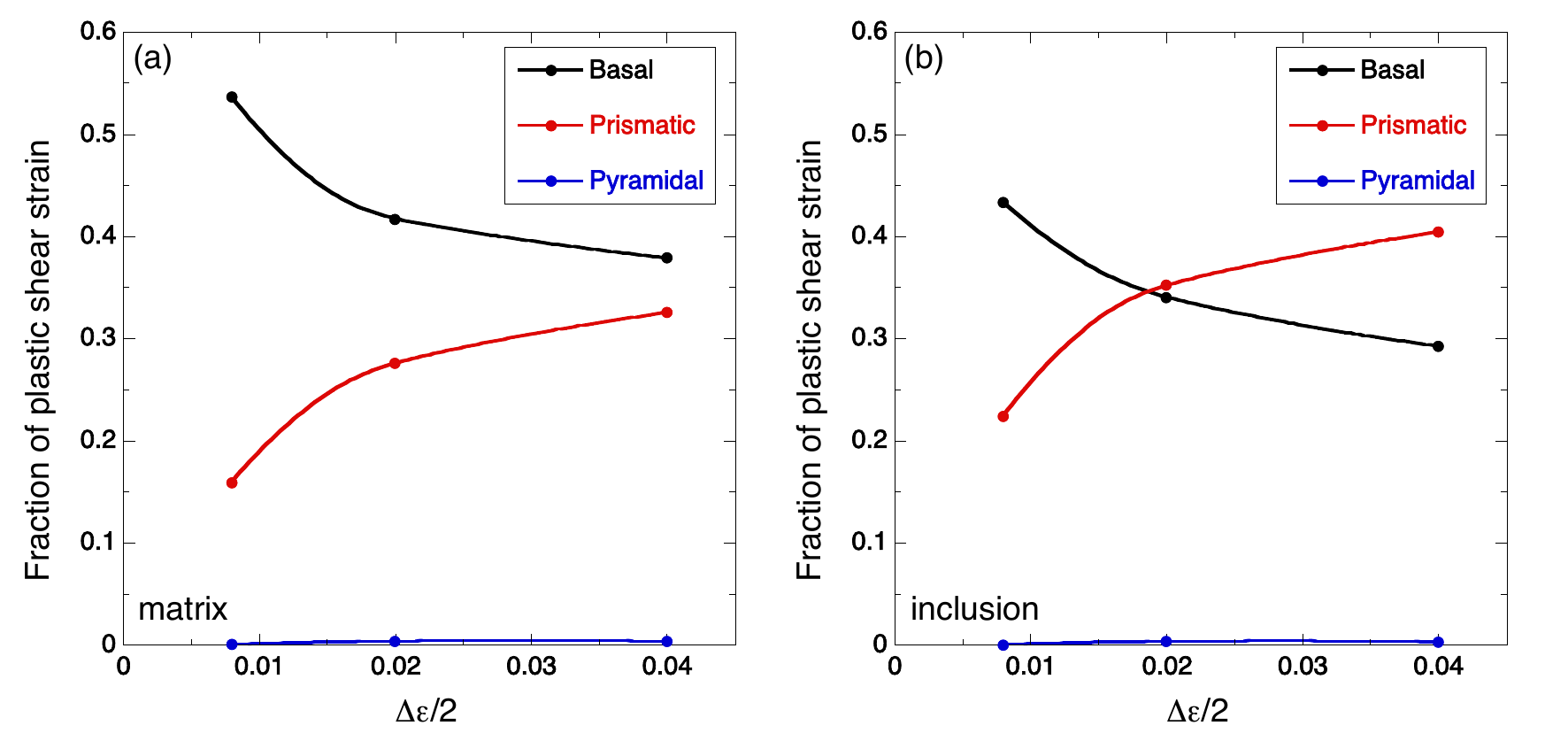}
\caption{Relative contribution of  basal, prismatic and pyramidal slip to the plastic shear strain during one fatigue cycle as a function of the applied cyclic strain semi-amplitude  $\Delta\protect\epsilon/2$. (a) Matrix region of the RVE. (b) Inclusion regions of the RVE.}
\label{fig:ShearStrain}
\end{figure}

The differences in the hardening rates in tension and compression between the matrix and inclusion regions can be understood from the activities of the different deformation mechanisms in each region during one fatigue cycle. The relative contribution of  basal, prismatic and pyramidal slip to the plastic shear strain during one fatigue cycle is plotted as  a function of the applied cyclic strain semi-amplitude,  $\Delta\protect\epsilon/2$ in Figs. \ref{fig:ShearStrain}a and b, for the matrix and inclusion regions of the RVE, respectively. At low cyclic strain amplitudes (and, thus, small applied stress), basal slip is the dominant deformation mechanism in the matrix and in the inclusions because the stresses are low and is more difficult to activate prismatic slip, which requires higher stresses. The differences in the relative contributions of basal and prismatic slip  between the matrix and the inclusions reflect the strong texture in the inclusions which are not suitably oriented for basal slip. The fraction of plastic strain accommodated by prismatic slip increased in both regions with the $\Delta\epsilon/2$ because of the higher stresses and, in fact, prismatic slip becomes dominant in the inclusions when $\Delta\epsilon/2$ = 4\%. Overall, the fraction of shear strain accommodated by prismatic slip in the inclusions was higher that that in the matrix, and this difference explained the higher strain hardening rate during the tensile part of the fatigue cycle in the inclusions (Fig.  \ref{fig:localSE}) 

\begin{figure}[tbp]
\centering
\includegraphics[width = 1.0\textwidth]{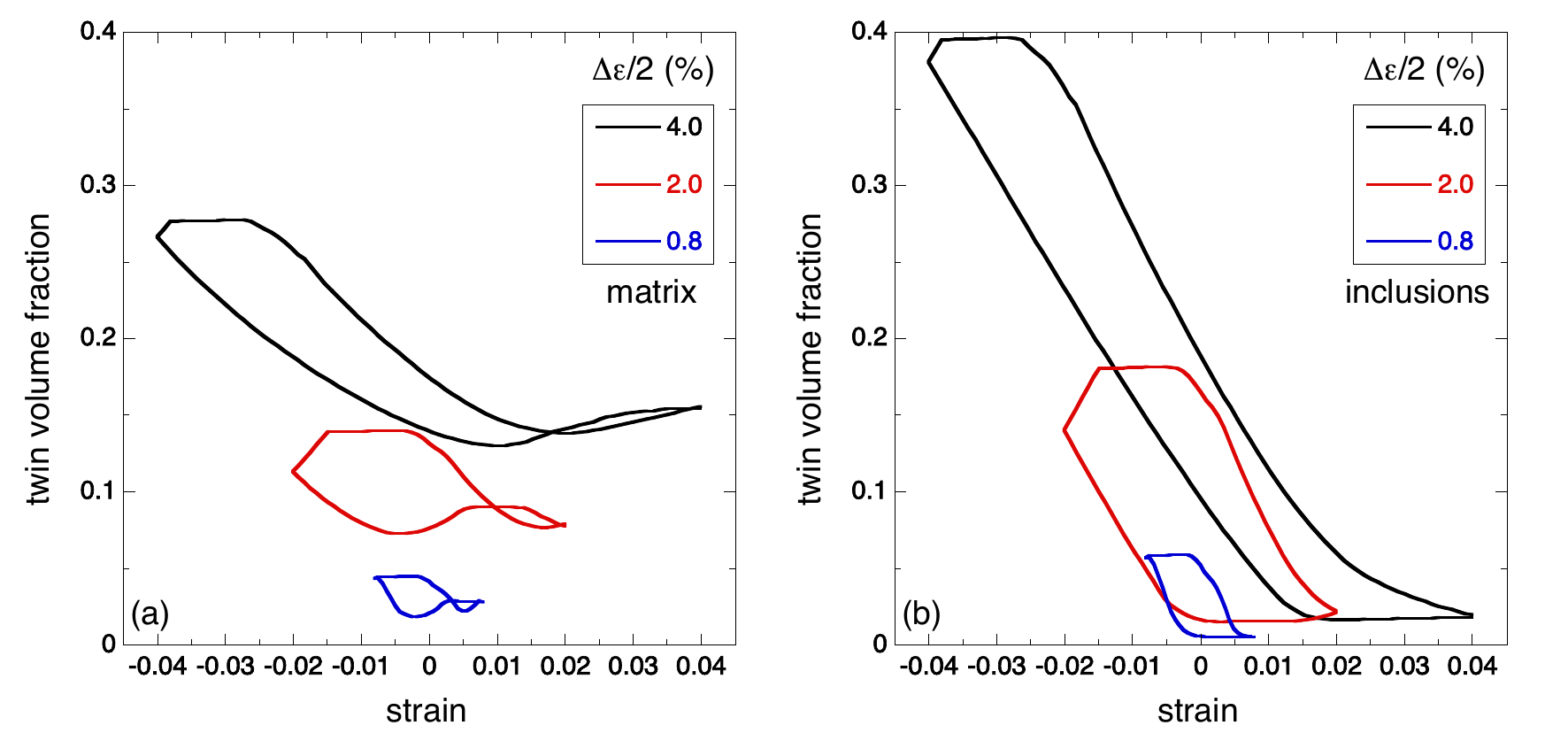}
\caption{Evolution of the twin volume fraction during one fatigue cycle for different cyclic strain amplitudes. (a) Matrix regions of the RVE. (b) Inclusion regions of the RVE.}
\label{fig:TwinEvolution}
\end{figure}

The results of the simulations in Fig. \ref{fig:ShearStrain} also show that pyramidal slip was not active during cyclic deformation and, thus, deformation of the Mg grains along the $c$ was accommodated by means of twinning. The evolution of the twin volume fraction in the matrix and  inclusion regions of the RVE during each fatigue cycle is plotted in Figs. \ref{fig:TwinEvolution}a and b, respectively,  for different values of the applied cyclic strain amplitude. The twin volume fraction increases rapidly during the compressive part of the fatigue cycle, particularly in the inclusions while detwinning is activated during the tensile part of the fatigue cycle. Full detwinning occurs in each fatigue cycle in the inclusion regions while some residual twinning is found in the matrix and the magnitude of this residual twin volume fraction increased with the applied cyclic strain amplitude. Full detwinning of the inclusions and partial detwinning of the matrix regions is also related to the differences in texture between both regions.

\subsection{Prediction of fatigue life}

Current strategies to predict the fatigue life of polycrystalline materials are based on the use of Fatigue Indicator Parameters (FIPs) \citep{Shenoy2007}. The FIPs are are associated to the main driving force that controls crack formation and can obtained from the evolution of mechanical fields and internal variables at the local level within the RVE in each fatigue cycle \citep{Segurado2018}. The most common ones are based exclusively on the local plastic strain fields such as the plastic shear strain accumulated in one fatigue cycle \citep{McDowell2003, Manonukul2004}. It should be noted that most of the fatigue life during low-cycle fatigue is dominated by the growth of small in cracks. Nevertheless, most of models to predict the fatigue life based on computational homogenization assume that the fatigue life can be predicted from FIPs obtained from RVEs without cracks \citep{Shenoy2007, Segurado2018}. This hypothesis is supported by the fact that many FIPs, such as the accumulated plastic strain or the energy dissipated in each fatigue, are good indicators of the driving force for both fatigue crack initiation and fatigue crack propagation of small cracks. 

In the case of a hcp Mg crystal with 3 basal, 3 prismatic and 12 pyramidal slip systems, the plastic shear strain accumulated in the slip system $\alpha$ of the voxel $V$ in each fatigue cycle can be determined as

\begin{equation}
\label{FIP_gamma}
\Delta\gamma^\alpha(V)= \int_{cyc} |\dot{\gamma}^\alpha(V)) | \;\mathrm{d}t
\end{equation}

\noindent where the contribution of twinning/detwinning to the plastic shear strain has not been included because this is a reversible process that does not contribute to the accumulation of damage. It might be that twinning-detwinning influences the nucleation of cracks during fatigue but this phenomenon is not clearly established. For instance, \cite{FWang2014} reported the mechanisms of microcrack initiation during fatigue of a Mg-G-Y alloy. They found that microcracks were nucleated at grain boundaries at high strain amplitudes and along persistent slip bands at low strain amplitudes. Twin boundaries were not identified as loci for fatigue crack initiation.

\begin{figure}[tbp]
\centering
\includegraphics[width = 0.6\textwidth]{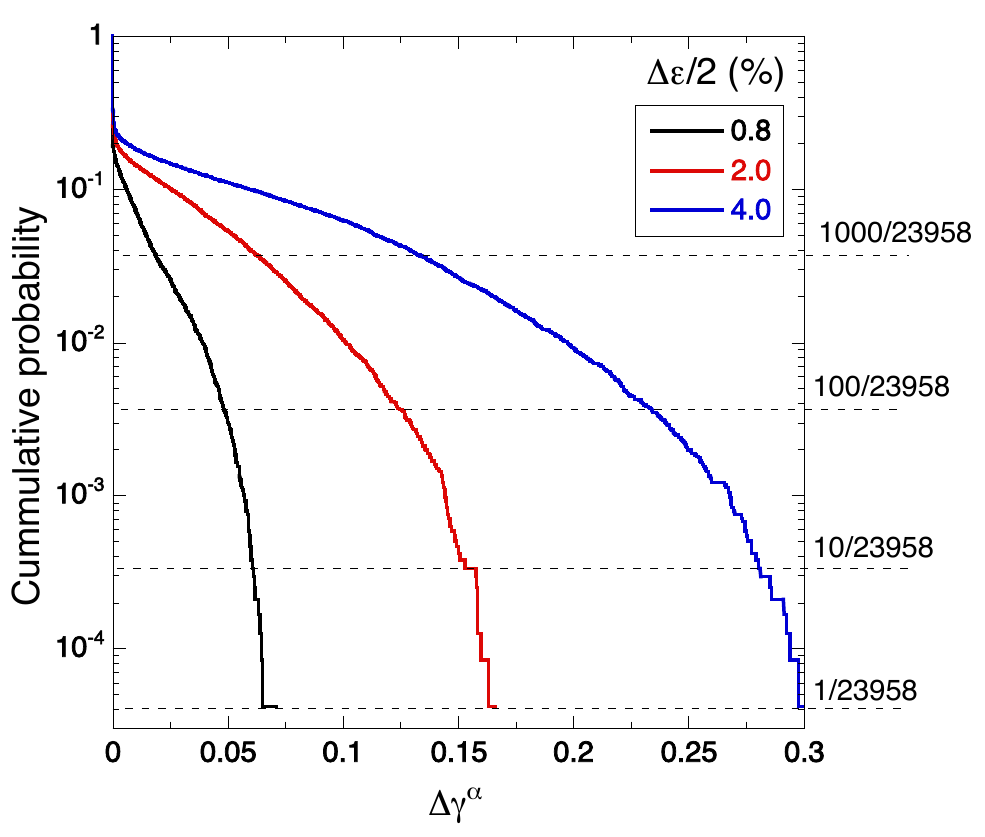}
\caption{Cumulative probability of $\Delta\gamma^\alpha$ in the RVE of the Mg alloy subjected to fatigue deformation at $\Delta\protect\epsilon/2$ = 0.8\%, 2.0\% and 4.0\%.}
\label{fig:CP}
\end{figure}

Obviously, $\Delta\gamma^\alpha(V)$ varies throughout the microstructure of the RVE and fatigue damage will be nucleated in the voxel with the highest accumulated plastic shear strain in stable fatigue cycles, which stands for the FIP. Nevertheless, it has been noticed that the maximum value of FIP may depend on the details of the finite element discretization. Various of averaging approaches were proposed to avoid spurious stress concentrations and minimize mesh size effects \citep{Castelluccio2015}. One possibility is to determine the FIP by volume averaging over a region representative of the crack incubation zone \citep{Shenoy2007, Castelluccio2015, Cruzado2018} but the mesh dependency can also be partially removed by extracting the FIP from \emph{the statistical distribution} of $\Delta\gamma^\alpha(V)$ throughout the RVE. There are 23958 slip systems in the RVE taking into account there are 1331 integration points and 18 slip systems (3 basal, 3 prismatic and 12 pyramidal) in each integration point. The cumulative probability of $\Delta\gamma^\alpha$ is plotted in Fig. \ref{fig:CP} for the RVE deformed at $\Delta\protect\epsilon/2$ = 0.8\%, 2.0\% and 4.0\%. The highest plastic shear strain accumulated in one slip system in each fatigue cycle, $\Delta\gamma^\alpha_1$, corresponds to a cumulative probability of 1/23958 while the values of $\Delta\gamma^\alpha$ which are attained in 10 ($\Delta\gamma^\alpha_{10}$), 100 ($\Delta\gamma^\alpha_{100}$) and 1000 ($\Delta\gamma^\alpha_{1000}$) slips systems of the RVE during one fatigue cycle are given by the  corresponding cumulative probabilities, as shown in Fig. \ref{fig:CP}. The highest values of $\Delta\gamma^\alpha$ were found within the matrix regions of the RVE.

The plastic shear strain accumulated that is attained in $s$ = 1, 10, 100 or 1000 slip systems during one fatigue cycle can be used to estimate the fatigue life according to

\begin{equation}
\label{FL}
N = \Delta\gamma^c / \Delta\gamma^\alpha_s
\end{equation}

\noindent where $\Delta\gamma^c$ is a model parameter whose value depends on the $\Delta\gamma^\alpha_s$ chosen as the FIP. The experimental results of the fatigue life in Fig. \ref{fig:EXP_fatigueLife} can be fitted to eq. \eqref{FL} using different values of $\Delta\gamma^\alpha_s$ for each FIP, which can be found in Table \ref{tab:gammaC_fitted}. They were obtained by the least squares fitting of the experimental results to eq. \eqref{FL}. The four FIPs used in eq. \eqref{FL} to estimate the fatigue life led to reasonable predictions and none of them can be selected as the best one because of the experimental scatter and the limited number of tests.  These results seem to indicate that the fatigue life of Mg alloys containing rare earths is controlled by the localization of plastic strain in each fatigue cycle. Due to the limited texture of these alloys, basal slip is the dominant deformation mechanism and, thus, controls the nucleation of fatigue cracks during cyclic deformation. The current study explicitly addresses the influence of the crystallographic texture on the fatigue life of Mg alloys and could be helpful to design Mg alloys with longer fatigue life.

\begin{figure}[tbp]
\centering
\includegraphics[width = 0.6\textwidth]{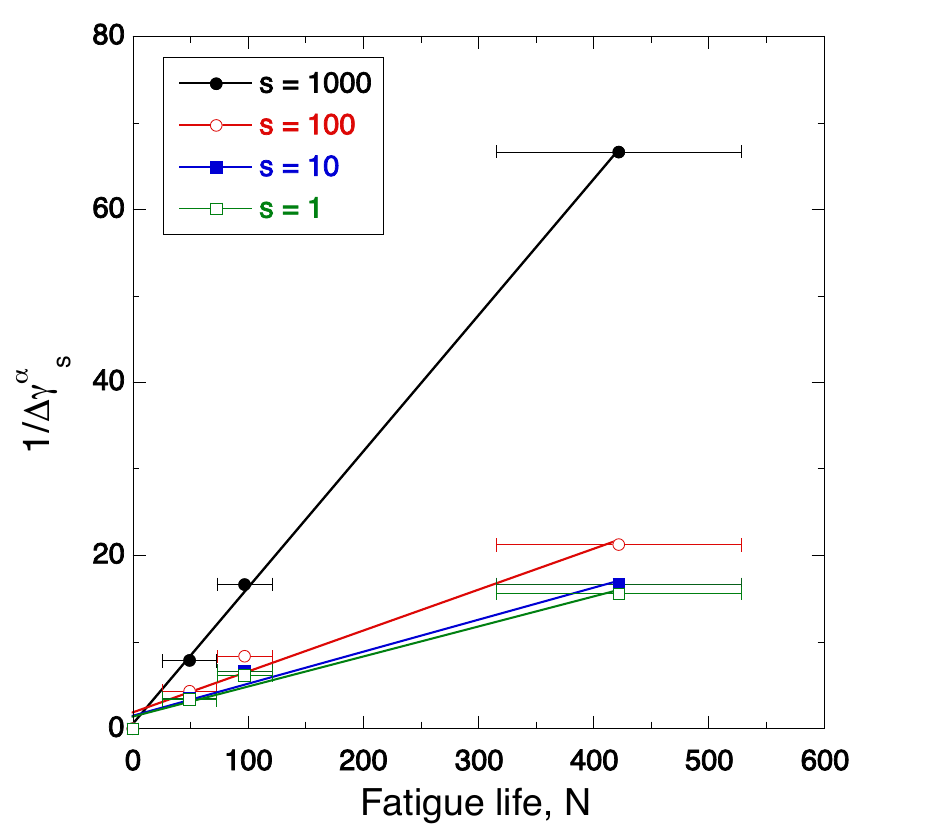}
\caption{Experimental results and estimations of the fatigue life from eq. \eqref{FL} for different $\Delta\gamma^\alpha_s$ FIPs. The dots of the experimental results stand for the average fatigue life while the error bars go from the minimum to the maximum fatigue life for each $\Delta\epsilon/2$.}
\label{fig:ES}
\end{figure}

\begin{table}[tbp]
\centering
\begin{tabular}{|c|c|c|c|c|}
\hline
  Fatigue indicator parameters & $\Delta\gamma^c$   \\ 
\hline
  $\Delta\gamma^\alpha_1$ &  28.0\\ 
  $\Delta\gamma^\alpha_{10}$ &  26.0\\ 
  $\Delta\gamma^\alpha_{100}$ &  20.5\\ 
  $\Delta\gamma^\alpha_{1000}$ &  6.35\\ 
\hline
\end{tabular}%
\caption{Critical values of the different fatigue indicator parameters to predict the fatigue life of the extruded Mg - 1Mn - 0.5Nd (wt. \%) alloy.}
\label{tab:gammaC_fitted}
\end{table}

\section{Conclusions}
The fatigue behavior under fully-reversed cyclic deformation was studied in an extruded Mg-1Mn-0.5Nd alloy. An advanced CP-FEM model was developed by considering important deformation mechanisms, i.e. basal slip, prismatic slip, and pyramidal slip, twinning and detwinning and back stress. At the same time, the computational homogenization was carried out using a 3D representative volume element. Although the pole figure of extruded Mg-1Mn-0.5Nd shew a typical extrusion texture, the magnitude of the texture is much smaller than the typical texture of Mg alloys without rare earth elements. Thus, the current paper reported the first detailed experiment-simulation-coupled study of the fatigue deformation mechanisms of Mg-RE alloy with weak texture. The main conclusion of our paper was the following:
\begin{itemize}
	\item Cyclic hardening under compression and cyclic softening under tension was observed due to the presence of twinning and detwinning. Moreover, the twin volume fraction of the broken samples depended on whether the sample was broken in tension or compression, indicating that twining-detwinning occurs during the whole fatigue life.
	\item The cyclic deformation was mainly accommodated by basal slip, prismatic slip and continuous twinning-detwinning. As full detwinning occurred in the inclusion regions but not in the matrix regions, texture influences fatigue life explicitly.	
	\item For the first time the fatigue life of the Mg-RE alloy was predicted from a fatigue indicator parameter based on the accumulated shear in a stable hysteresis loop.
\end{itemize}
 
\section{Acknowledgements}

This investigation was supported by the European Union Horizon 2020 research and innovation programme (Marie Sklodowska-Curie Individual Fellowships, Grant Agreement 795658) and the Comunidad de Madrid Talento-Mod1 programme (Grant Agreement PR-00096). Additional support from the European Research Council under the European Union's Horizon 2020 research and innovation programme (Advanced Grant VIRMETAL, grant agreement No. 669141) and by the HexaGB project of the Spanish Ministry of Science (reference RTI2018-098245) is also gratefully acknowledged. The authors thank to Dr. M. T. P\'erez-Prado who supplied the Mg alloy used in this investigation.

\bibliographystyle{model2-names}
\bibliography{literature}

\end{document}